\documentclass[a4paper,11pt]{article}

\usepackage{cite}
\usepackage{jheppub}

\usepackage{atlasphysics}
\usepackage{subfigure}

\usepackage{preprintcover} 
\PreprintCoverPaperTitle{ Measurement of differential production cross-sections for a $Z$ boson in association with $b$-jets in 7~TeV proton-proton collisions with the ATLAS detector}
\PreprintIdNumber{CERN-PH-EP-2014-118}
\PreprintCoverAbstract{Measurements of differential production cross-sections of a $Z$ boson in association with $b$-jets in $pp$ collisions at $\sqrt{s}=7$~TeV are reported. The data analysed correspond to an integrated luminosity of 4.6~fb$^{-1}$ recorded with the ATLAS detector at the Large Hadron Collider. Particle-level cross-sections are determined for events with a $Z$ boson decaying into an electron or muon pair, and containing $b$-jets. For events with at least one $b$-jet, the cross-section is presented as a function of the $Z$ boson transverse momentum and rapidity, together with the inclusive $b$-jet cross-section as a function of $b$-jet transverse momentum, rapidity and angular separations between the $b$-jet and the $Z$ boson. For events with at least two $b$-jets, the cross-section is determined as a function of the invariant mass and angular separation of the two highest transverse momentum $b$-jets, and as a function of the $Z$ boson transverse momentum and rapidity. Results are compared to leading-order and next-to-leading-order perturbative QCD calculations.}
\PreprintJournalName{JHEP}

\def\pythia{{\sc Pythia}}
\def\herwig{{\sc Herwig}}
\def\jimmy{{\sc Jimmy}}
\def\alpgen{{\sc Alpgen}}
\def\sherpa{{\sc Sherpa}}
\def\mcfm{{\sc mcfm}}
\def\auet{{AUET2}}
\def\ambt{{AMBT2b}}
\def\cteqsol{{CTEQ6L1}}
\def\cteqfl{{CTEQ5L}}
\def\ctten{{CT10}}
\def\mcnlo{{\sc mc$@$nlo}}
\def\amcnlo{a{\sc mc$@$nlo}}
\def\acermc{{\sc AcerMC}}
\def\powheg{{\sc Powheg}}
\def\evtgen{{\sc EvtGen}}

\def\geant{{\sc Geant4}}
\def\photos{{\sc Photos}}
\def\herwigpp{{\sc Herwig++}}
\def\onetag{{1-tag}}
\def\twotag{{2-tag}}
\def\Zb{{$Z+\ge 1$ $b$-jet}}
\def\Zbb{{$Z+\ge 2$ $b$-jets}}
\def\sigZbjet{{$\sigma(Zb)\times N_{b\text{-jet}}$ }}
\def\sigZbjetstar{{$\sigma^{*}(Zb)\times N_{b\text{-jet}}$}}
\def\sigZb{{$\sigma(Zb)$}}
\def\sigZbb{{$\sigma(Zbb)$}}
\def\dR{{$\Delta {\cal R}$}}
\def\dRZb{{$\Delta R(Z,b)$}}
\def\absy{{$|y|$}}
\def\mbb{{$m(b,b)$}}
\def\dRbb{{$\Delta R(b,b)$}}

\title{
\boldmath Measurement of differential production cross-sections for a $Z$ boson in association with $b$-jets in 7~\tev\ proton-proton collisions with the ATLAS detector}

\author{
The ATLAS Collaboration
}

\abstract{
Measurements of differential production cross-sections of a $Z$ boson in association with $b$-jets 
in $pp$ collisions at $\sqrt{s}=7$~TeV are reported. The data analysed correspond to an 
integrated luminosity of 4.6~fb$^{-1}$ recorded with the ATLAS detector at the Large 
Hadron Collider. Particle-level cross-sections are determined for events with
a $Z$ boson decaying into an electron or muon pair, and containing $b$-jets. 
For events with at least one $b$-jet, the cross-section is presented as a function of the 
$Z$ boson transverse momentum and rapidity, together with the inclusive $b$-jet cross-section as a 
function of $b$-jet transverse momentum, rapidity and angular separations between the 
$b$-jet and the $Z$ boson. For events with at least two $b$-jets, the cross-section 
is determined as a function of the invariant mass and angular separation 
of the two highest transverse momentum $b$-jets, and as a function of the $Z$ boson transverse momentum and 
rapidity. Results are compared to leading-order and next-to-leading-order perturbative QCD calculations.}

\begin{document} 
\maketitle
\flushbottom

\section{Introduction}
\label{sec:Introduction}
The production of a $Z$ boson (using $Z$ as shorthand for $Z/\gamma^*$) decaying to 
electrons or muons provides a clear experimental signature at a hadron collider, 
which can be used as a probe of the underlying collision processes.
Such events provide an opportunity for the study of associated heavy flavour 
production and dynamics, which can be experimentally identified by reconstructing displaced 
decay vertices associated with the relatively long lifetimes of $b$-hadrons.
Predictions for heavy flavour production typically suffer from larger theoretical 
uncertainties than those for the more inclusive $Z$+jets processes, and measurements of 
$Z$ boson production in association with $b$-jets can therefore provide important 
experimental constraints to improve the theoretical description of this process.
The $Z$+$b$-jets signal is also an important background
to $ZH$ associated Higgs boson production with $H\rightarrow b\bar{b}$, 
as well as for potential signatures of physics beyond the Standard Model containing leptons 
and $b$-jets in the final state. 

Two schemes are generally employed in perturbative QCD (pQCD) calculations containing 
heavy flavour quarks.
One is the four-flavour number scheme (4FNS), which only considers parton densities of gluons 
and of the first two quark generations in the proton.
The other is the five-flavour number scheme (5FNS), which allows a $b$-quark density in the 
initial state 
and raises the prospect that measurements 
of heavy flavour production could constrain the $b$-quark parton density function (PDF) of the proton.
In a calculation to all orders, the 4FNS and 5FNS methods must give identical results; 
however, at a given order differences can occur between the two. 
A recent discussion on the status of theoretical calculations and the advantages and disadvantages of 
the different flavour number schemes can be found in Ref.~\cite{binitMaltoni}. 

\begin{figure}[bp]
\begin{center}
\subfigure[ ]{
  \includegraphics[width= 0.3\columnwidth]{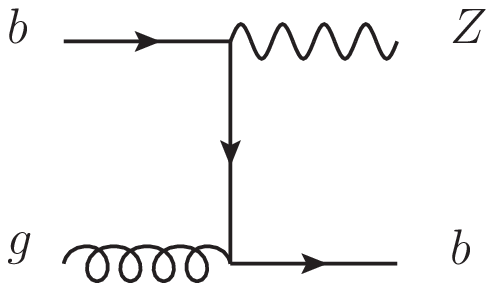}
  \label{fig:feynman1}
}
\subfigure[ ]{
  \includegraphics[width= 0.3\columnwidth]{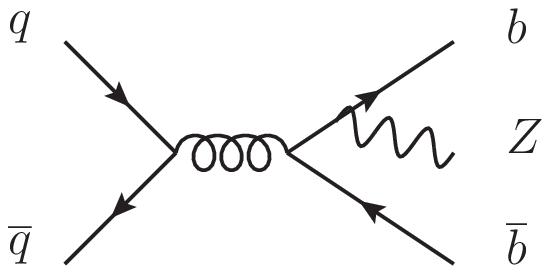}
  \label{fig:feynman2}
}
\subfigure[ ]{
  \includegraphics[width= 0.3\columnwidth]{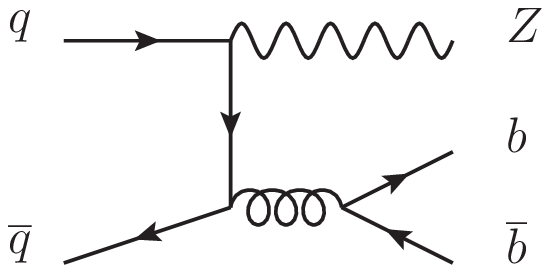}
  \label{fig:feynman3}
}
\end{center}
\caption{Leading order Feynman diagrams contributing to $Z$+$b$-jets production. Process \ref{fig:feynman1} is only present in a 5FNS calculation, while \ref{fig:feynman2} and \ref{fig:feynman3} are present in both the 4FNS and 5FNS calculations.  
\label{fig:feynman}}
\end{figure}

Next-to-leading-order (NLO) matrix element calculations have been 
available for associated $Z$+$b$ and $Z$+\bbbar\ production at parton-level for a number of years 
\cite{MCFMZb,MCFMZbj,ZbbCordero}. The leading order (LO) Feynman diagrams shown
in Figure~\ref{fig:feynman} illustrate some of the contributing processes. 
Full particle-level predictions have existed at LO for some time, obtained by
 matching parton shower generators to LO multi-leg matrix elements in the 4FNS \cite{Alpgen, acermc}, 
5FNS \cite{Sherpa}, or both~\cite{madgraph}. More recently, a full particle-level prediction for \Zbb\ at NLO in the 4FNS with matched parton shower has become available \cite{amcnlo_zbb,aMCNLO}. The same framework can also be used
to provide a full particle-level prediction for \Zb\ at NLO in the 5FNS. 
In this article data are compared with several theoretical predictions following different approaches.

Differential measurements of $Z$+$b$-jets production have been made in proton-antiproton collisions 
at \rts=1.96~\tev\ by the CDF and D0 experiments \cite{cdfZb,d0Zb} as well as inclusively 
in \rts=7~\tev\ proton-proton collisions at the LHC by the ATLAS and CMS experiments \cite{atlasZb, cmsZb}. 
The results presented in this paper significantly extend the scope
of the previous ATLAS measurement, which used around 36~\ipb of data recorded in 2010. The current analysis takes advantage of the full sample of \rts=7~\tev\ 
proton-proton collisions recorded in 2011, corresponding to an integrated luminosity 
of 4.6~\ifb, and uses improved methods for $b$-jet identification to cover 
a wider kinematic region. The larger data sample allows differential 
production cross-section measurements of a $Z$ boson with $b$-jets at 
the LHC. These complement the recently reported results of associated production of a 
$Z$ boson with two $b$-hadrons at \rts=7~\tev\ by CMS \cite{cmsZBB}.

A total of 12 differential cross-sections are presented here, covering a variety of $Z$ boson and $b$-jet
kinematics and angular variables sensitive to different aspects of the theoretical predictions, 
as listed in Table~\ref{tab:intro:diffZbb}.
All cross-sections include the $Z$ boson branching fraction, ${\mathrm{ Br}}(Z\rightarrow \ell^{+}\ell^{-})$, 
where  $\ell$ is a single lepton flavour, and are reported in a restricted fiducial region, defined using 
particle-level quantities, detailed in Section~\ref{sec:Unfolding}, which are chosen to 
minimise extrapolation from the corresponding measured detector-level quantities.

The results are grouped according to different selections which give four integrated cross-section definitions: 
\begin{itemize}
\item \sigZb, the cross-section for events containing a $Z$ boson and one or more $b$-jets in the fiducial region;
\item \sigZbjet, the inclusive cross-section for all $b$-jets in the fiducial region in events with a $Z$ boson; 
\item \sigZbjetstar, similar to \sigZbjet, with the additional requirement that the dilepton system has transverse momentum, $\pt>20$~\gev, ensuring the $\phi(Z)$ 
coordinate\footnote{ATLAS uses a right-handed coordinate system, with the origin at the nominal interaction point, with the beam line defining the $z$ axis, the $x$-axis pointing towards the centre of the LHC ring, and the $y$-axis vertically up. The azimuthal angle, $\phi$, is defined in the transverse ($x-y$) plane, and the pseudo-rapidity is used instead of the polar angle: $\eta=-\ln \tan(\theta/2)$. Rapidity is defined in the usual way, $y = \ln[(E + p_z)/(E -p_z)]/2$.} (which is taken from the direction of the 
dilepton system) is well defined and not limited by detector resolution. This is necessary for the differential measurements of $\Delta \phi(Z,b)$ and hence $\Delta R(Z,b)$\footnote{Two measures of angular separation are used: $\Delta {\cal R}=\sqrt{\Delta\phi^2 + \Delta\eta^2}$, and $\Delta R=\sqrt{\Delta\phi^2 + \Delta y^2}$.}.
\item \sigZbb, the cross-section for events containing a $Z$ boson and two or more $b$-jets in the fiducial region. 
When there are more than two $b$-jets, quantities are calculated using the two highest \pt\ $b$-jets in the event.
\end{itemize}

This paper will cover the experimental apparatus, simulation and event selection 
 in Sections~\ref{sec:Detector},~\ref{sec:Simulation} and~\ref{sec:Selection}, followed by the description of the methods used to
determine backgrounds and extract the signal in Sections~\ref{sec:Background} and~\ref{sec:Signal}.
Conversion of the measured data to differential cross-sections and the details of the systematic 
uncertainties are covered in Sections~\ref{sec:Unfolding} and~\ref{sec:Systematics}.
A number of theoretical predictions, detailed in Section~\ref{sec:Theory}, are compared to the data in Section~\ref{sec:Results}, before conclusions are drawn in Section~\ref{sec:Conclusions}.

\begin{table}[h]
\begin{center}
\begin{tabular}{|c|c|c|c|}
  \hline
  Variable & Definition & Range & Integrated \\
  & & & cross-section \\ 
  \hline
  $\pt(Z)$ & $Z$ boson transverse momentum & 0--500~\gev & \sigZb \\
  $|y(Z)|$ & $Z$ boson absolute rapidity & 0.0--2.5 & \sigZb \\
  \hline
  $\pt(b)$ & $b$-jet \pt & 20--500~\gev &  \sigZbjet  \\
  $|y(b)|$ & $b$-jet absolute rapidity & 0.0--2.4 & \sigZbjet  \\
  $y_{\text{boost}}(Z,b)$ & $|(y(Z) + y(b))|/2.0$ & 0.0--2.5 & \sigZbjet \\
  \hline
  $\Delta y(Z,b)$ & $|y(Z)-y(b)|$ & 0.0--5.0 & \sigZbjetstar \\
  $\Delta \phi(Z,b)$ & $|\phi(Z)-\phi(b)|$ & 0.0--$\pi$ & \sigZbjetstar \\
  $\Delta R(Z,b)$ & $\sqrt{\Delta \phi(Z,b)^{2} + \Delta y(Z,b)^{2}}$ & 0.0--6.0 & \sigZbjetstar \\
  \hline
  $\pt(Z)$ & $Z$ boson transverse momentum & 0--250~\gev & \sigZbb \\
  $|y(Z)|$ & $Z$ boson absolute rapidity & 0.0--2.5 & \sigZbb \\
  $m(b,b)$ & $bb$ invariant mass & 20--350~\gev & \sigZbb \\
  $\Delta R(b,b)$ & $bb$ angular separation & 0.4--5.0 & \sigZbb \\
  \hline
\end{tabular}
\caption{Definitions of variables for which differential production cross-sections are measured and the ranges over which those measurements are performed. The integral of each differential cross-section yields one of the four integrated cross-sections defined in the text.}
\label{tab:intro:diffZbb}
\end{center}
\end{table}

\section{The ATLAS experiment}
\label{sec:Detector}

The ATLAS experiment~\cite{atlas} is a multi-purpose particle detector with large solid
angle coverage around one of the interaction regions of the LHC.
It consists of an inner tracking detector surrounded by a superconducting
solenoid providing a 2~T axial magnetic field, followed by electromagnetic and hadronic 
calorimeters and a muon spectrometer with three superconducting toroid magnets. 
The inner detector (ID) is made up of a high-granularity silicon pixel detector, a silicon 
microstrip tracker, and a straw-tube transition radiation tracker. These provide measurements of charged 
particles in the region $|\eta|< 2.5$.
The calorimeter system covers $|\eta|< 4.9$ and utilises a variety of absorbing and sampling technologies.
For $|\eta|<3.2$, the electromagnetic (EM) calorimeters are based on high-granularity lead/liquid-argon (LAr), while the $3.1<|\eta|<4.9$ forward region uses copper/LAr.
Hadronic calorimetery is based on steel and scintillating tiles for $|\eta|<1.7$, copper/LAr for $1.5<|\eta|<3.2$, and tungsten/LAr for $3.1<|\eta|<4.9$.
The muon spectrometer (MS) comprises resistive plate and thin gap trigger chambers covering $|\eta|<2.4$,  and high-precision drift tubes and cathode strip tracking chambers, covering $|\eta|<2.7$. 
ATLAS uses a three-level trigger system to select potentially interesting collisions. 
The Level-1 trigger is hardware based, and uses a subset of detector information to reduce the event rate to at most 75~kHz. Two software-based trigger levels follow, which reduce the event rate to about
300~Hz, for offline analysis.

\section{Simulated event samples}
\label{sec:Simulation}
The Monte Carlo (MC) simulations of proton-proton collisions and the expected response of the ATLAS detector to simulated particles are used in three ways in this analysis: first, to estimate signal and background 
contributions to the selected data sample; second, to determine correction factors for detector effects and acceptance when calculating 
particle-level cross-sections; and finally to estimate systematic uncertainties.

Inclusive $Z(\rightarrow \ell\ell)$ events, produced in associations with both light and heavy flavour jets, are simulated using \alpgen\ 
2.13 \cite{Alpgen} interfaced to \herwig\ 6.520 \cite{Herwig} to model the parton shower and hadronisation, and 
\jimmy\ 4.31 \cite{Jimmy} to model the underlying event and multi-parton interactions (MPI). \alpgen\ produces matrix 
elements with up to five partons using a LO multi-legged approach; these are matched to final state jets using 
the MLM method \cite{MLMmatching} to remove overlaps in phase-space between events containing jets produced in the matrix element and jets produced in the parton shower. Samples are generated with the \cteqsol\ \cite{cteqsol} PDF set and the \auet\ tuning of parameters \cite{AUETtwo} for the description of the non-perturbative component of the generated events. 
In addition, overlaps between samples with heavy-flavour quarks originating from the matrix element and from the parton shower are removed.
Events containing $b$-quarks are reweighted after hadronisation to reproduce $b$-hadron decay particle multiplicities predicted by the \evtgen\ package \cite{Evtgen}, to correct mismodelling found in the decay tables of the \herwig\ generator version used. 
Alternative $Z(\rightarrow \ell\ell)$ samples used for 
systematic cross-checks are generated with \sherpa\ 1.4.1 \cite{Sherpa}. This generator is based on a 
multi-leg matrix element calculation using the \ctten\ \cite{ct10} PDF set and matched to the parton shower using the CKKW prescription~\cite{CKKW}.

Backgrounds from \ttbar, single top quark production in the $s$-channel, $W+t$ production, and diboson processes are simulated using \mcnlo\ 4.01 \cite{mcnlo} interfaced to 
\herwig\ and \jimmy\ using the \ctten\ PDF set. Single top quark production in the $t$-channel is generated with \acermc\ 3.7 \cite{acermc}
interfaced to \pythia\ 6.425 \cite{pythia} using the \cteqsol\ PDF set.
Corrections to \herwig\ $b$-hadron decay tables using \evtgen\ 
are made for both \ttbar\ and $ZZ(\rightarrow\bbbar\ell\ell)$ events which are the dominant backgrounds containing real 
$b$-jets. Samples of $W(\rightarrow \ell\nu)$ events
are generated using \alpgen\ interfaced to \herwig\ and \jimmy\ in an identical
configuration to that used for $Z(\rightarrow\ell\ell)$+jets events described above. An alternative \ttbar\ sample used for systematic cross-checks is generated
with \powheg\ \cite{powheg} interfaced to \pythia\ using the \ctten\ PDF set.

The total cross-sections of the $W$, $Z$\ and \ttbar\ simulated samples are normalised to NNLO predictions~\cite{fewz, ttbar_xs}, while other backgrounds
are normalised to NLO predictions~\cite{diboson_xs, singletop_xs}. All samples are overlaid with minimum bias interactions, generated with \pythia\ 6.425
 using the \cteqsol\ PDF set and \ambt\ tune \cite{AMBTtwoB}, to simulate multiple interactions per bunch crossing (pile-up) such that 
the distribution of the average number of interactions observed in 2011 $pp$ collision data, with mean value of 9.1, is accurately reproduced. Furthermore, the samples are weighted such that the $z$ distribution of reconstructed $pp$ interaction vertices matches the
distribution observed in data. The ATLAS detector response is modelled using the 
\geant\ toolkit \cite{geant,atlassim}, and event reconstruction similar to that used for data is performed.

\section{Event selection}
\label{sec:Selection}
The data analysed were collected by the ATLAS detector in 2011 during stable $pp$ collisions at \rts=7~\tev\ 
when all components of the ATLAS detector were fully functioning. Dielectron candidate events were selected
with a trigger requiring two electrons with $\pt>12$~\gev. Dimuon
candidate events were selected with a trigger requiring a single muon with $\pt>18$~\gev. 
An integrated luminosity of 4.58$\pm$0.08~\ifb\ \cite{lumipaper} was taken with these triggers.

The primary interaction vertex (PV) is defined as the vertex with highest $\sum\pt^2$ of ID tracks with 
$\pt>0.4$~\gev\ associated to it. Candidate
events are required to have at least three such associated tracks.

Electron candidates are reconstructed by associating a cluster of energy deposits in the EM calorimeter to a well 
reconstructed ID track, and are required to have $\et>20$~\gev\
and $|\eta|<2.47$, excluding the region $1.37<|\eta|<1.52$ where the transition between the barrel and endcap of the EM calorimeter occurs.
Candidates are required to pass a `medium' quality requirement based on analysis of various 
cluster properties and the associated ID track~\cite{elecperf}. Muon candidates are reconstructed by associating well 
identified ID tracks to MS tracks \cite{muonperf}. Candidates are required to have $\pt>20$~\gev\ 
and $|\eta|<2.4$. Selections on the transverse energy (transverse momentum) of electron
(muon) candidates are chosen to ensure the trigger is fully efficient.

To ensure that lepton candidates originate from the PV and to suppress those candidates originating from heavy flavour decays,
ID tracks associated to lepton candidates are required to have an absolute 
longitudinal impact parameter with respect to the PV, $|z_{0}|$, less than 1~mm  and absolute transverse 
impact parameter, $|d_{0}|$, no larger than ten (three) times its measured uncertainty for electron (muon) 
 candidates. Muon candidates are additionally required to be isolated from local track activity by rejecting 
candidates where the summed transverse momenta of additional ID tracks within \dR~$=0.2$ from the muon candidate
is larger than 10\% of the transverse momentum of the candidate itself. No additional isolation requirement is 
applied to electron candidates as the quality requirement and impact parameter selections already sufficiently 
reduce the contribution from jets misidentified as electrons in the calorimeter.

Selection efficiencies of lepton candidates as well as their energy resolution and momentum resolution 
are adjusted in simulation to match those observed in $Z\rightarrow\ell\ell$ events in 
data \cite{elecperf,muonperf}. The lepton energy scales and momentum scales are 
calibrated based on a comparison of the position of the $Z$ boson mass peak in data and simulation. 
Events with exactly two lepton candidates of same flavour and opposite measured 
charge are kept for further analysis, provided the invariant  mass of those leptons, 
$m_{\ell\ell}$, falls in the range $76<m_{\ell\ell}<106$~\gev.

Jets are reconstructed from topological energy clusters in the calorimeter \cite{topoclus} using the 
anti-$k_{t}$ algorithm \cite{aktalg, akt2} with radius parameter $R=0.4$. The jet energy is calibrated 
as a function of \pt\ and $\eta$ using MC simulation after correcting first for the energy offset 
due to pile-up activity in the calorimeter, and then redefining the origin of the jet to be the event PV. 
A residual in situ correction determined from $Z$+jet and $\gamma$+jet control samples is 
applied to jets in data to account for remaining differences in calorimeter response between data and 
simulation \cite{jes2011}. Jets from pile-up interactions are suppressed by requiring that ID tracks associated 
to the PV contribute at least 75\% of the total scalar sum of transverse momenta from all ID tracks within \dR~$=0.4$ from the 
jet centroid. Calibrated reconstructed jets are required to have $\pt>20$~\gev, $|y|<2.4$ 
and any jet within \dR~$=0.5$ of a selected lepton candidate is removed.

Jets containing $b$-hadrons are identified using a neural network (NN) algorithm, 
MV1~\cite{btag}. The MV1 algorithm takes as inputs the results of lower-level likelihood and 
NN based $b$-tagging algorithms, which in turn take both secondary vertex
kinematics and impact parameter information with respect to the PV as inputs,
obtained from analysing ID tracks within \dR~$=0.4$ from the jet centroid. 
The MV1 variable lies in the range [0,1] with a value closer to 
unity denoting a higher probability for the jet to be a $b$-jet. Reconstructed
 $b$-jet candidates are selected for the analysis when their MV1 output is greater 
than the value corresponding to a 75\% average $b$-tagging efficiency in 
simulated \ttbar\ events.
In simulation, reconstructed jets are labelled as
 $b$-jets if they lie within \dR~$=0.3$ from one or more weakly decaying $b$-hadrons 
with $\pt>5$~\gev. Reconstructed jets not identified as $b$-jets are
considered as $c$-jets if they lie within \dR~$=0.3$ from any $c$-quark with 
$\pt>5$~\gev. All other jets are classified as `light-jets'. Tagging efficiencies 
in simulation are scaled to match those measured in data for all flavours as a
function of jet \pt\ (and $\eta$ for light-jets) using weights derived 
from control samples enriched in jets of each flavour \cite{btag, ctag, ltag}.

In each event, the  missing transverse momentum, \met, is also used to reject backgrounds which typically
contain high energy neutrinos, such as \ttbar. The \met\ is calculated by first forming the vector sum of 
all calibrated leptons and jets, along with any additional topological energy clusters not already associated 
to a reconstructed physics object. The magnitude of this sum in the transverse direction is a measure of the 
energy imbalance in the event, and is taken as the \met \cite{MET2011}.

Events used for further analysis are separated into two categories: 
those with at least one tagged jet, 
referred to as \onetag\ events; and those with at least two 
tagged jets, referred to as \twotag\ events, which is a subset of the 
\onetag\ sample.

\section{Background estimation and reduction}
\label{sec:Background}
Selected events in data contain the signal of interest as well as various background processes
 with either real or fake leptons and real or fake $b$-jets. By far the dominant contributions 
are $Z$+jets events where either a light-jet or $c$-jet has been misidentified as a $b$-jet. 
The amount of this background present in data is determined using fits to data as 
described in Section~\ref{sec:Signal}.

The next most important background arises from \ttbar\ events where both $W$ bosons decay to leptons. 
This background is estimated using simulated events normalised to the theoretically predicted cross-section.
The \ttbar\ background is suppressed by the requirements on $m_{\ell\ell}$, and its overall contribution to the event 
sample is small. However, it can be significant in some kinematic regions, particularly at higher jet \pt. 
To further reduce the \ttbar\ contamination, events are required to have \met~$< 70$~\gev. 
Figure~\ref{fig:etmiss} shows the \met\ distributions for signal and \ttbar\ simulations in 
\onetag\ and \twotag\ events after combining the electron and muon channels. 
The 70~\gev\ selection rejects 44.8\% (44.3\%) of the \ttbar\ background in \onetag\ (\twotag) events while remaining over 99\% efficient for signal events.

The total contribution to the final data sample from single top quark and diboson processes is estimated using samples of 
simulated events normalised to their theoretically predicted cross-sections.
Other electroweak processes such as $W$+jets and \Ztau\ events are found to have a negligible contribution in the selected phase space.

\begin{figure}[h]
\begin{center}
\subfigure[ ]{
  \includegraphics[width= 0.47\textwidth]{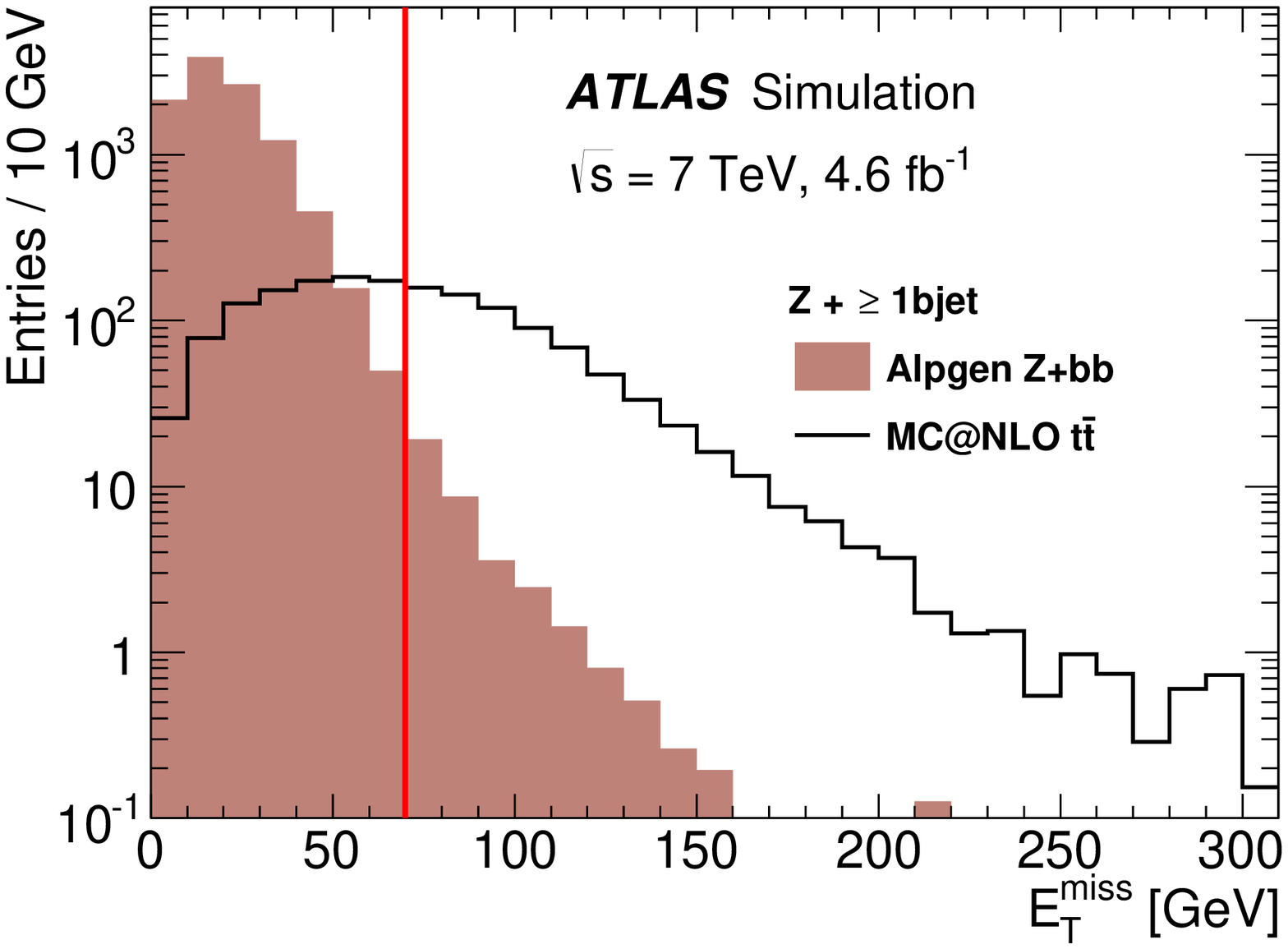}
  \label{fig:etmiss_Zb}
}
\subfigure[ ]{
  \includegraphics[width=0.47\textwidth]{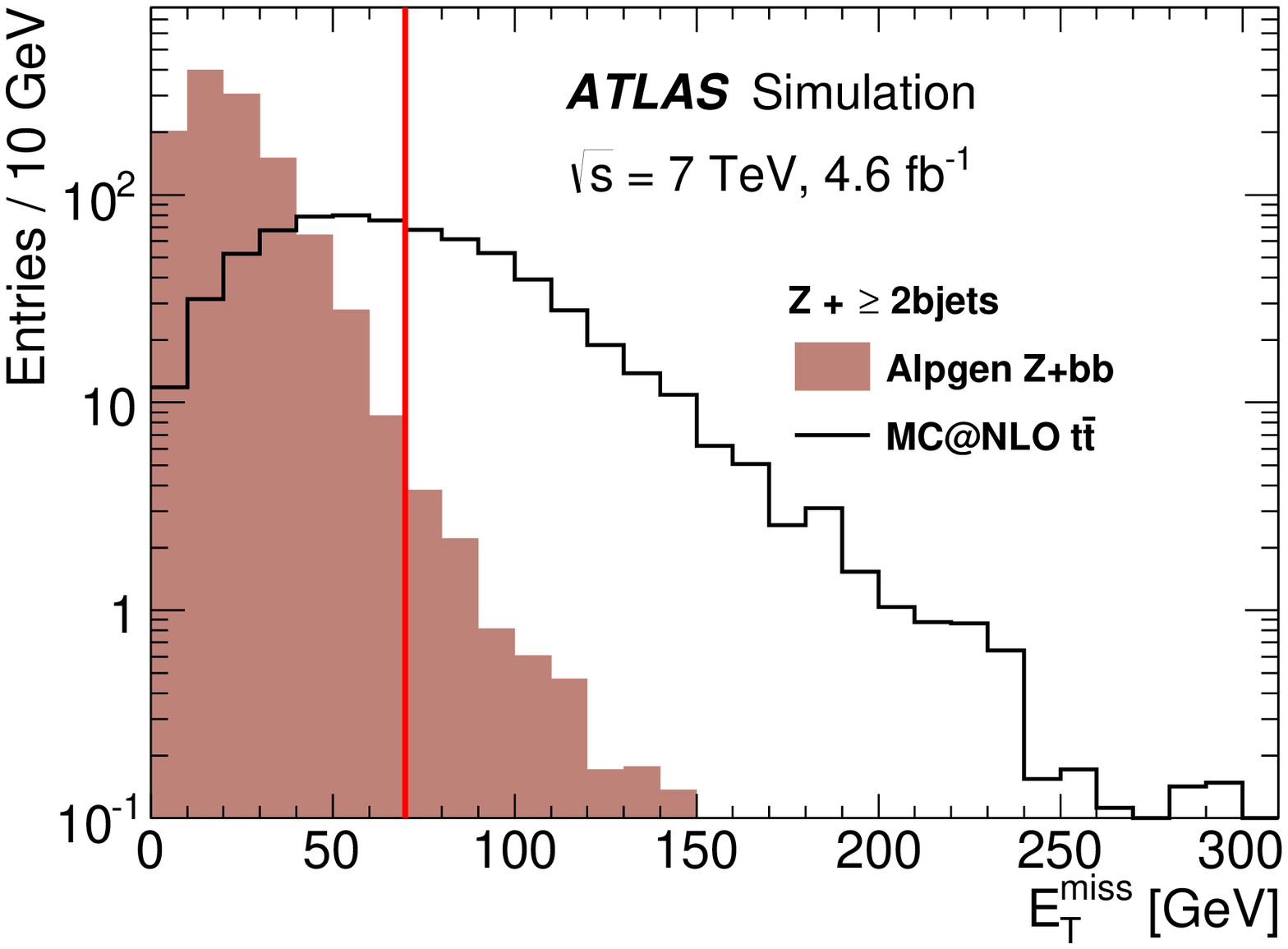}
  \label{fig:etmiss_Zbb}
}
\end{center}
\caption{Comparison of simulated \met\ distributions for (a) 1-tag events and (b) 2-tag events after all other signal selection criteria are applied, normalised to the expected yields in the data sample. 
The shaded distributions are signal \alpgen+\herwig+\jimmy\ events, and the  open distributions 
are selected \ttbar\ events. The vertical line shows the selection applied to the analysis sample 
to reject \ttbar\ events while keeping signal events. 
\label{fig:etmiss}}
\end{figure}

Background contributions from multijet events are estimated using data-driven techniques separately 
in the electron and muon channels for both \onetag\ and \twotag\ events. Multijet-enriched control 
regions are used to derive the expected shape of this background in the $m_{\ell\ell}$ variable.
These control regions drop the $b$-tagging requirement, and use an extended range $50<m_{\ell\ell}<200$~\gev\  
in order to maximise the available sample size. Studies found that no bias was introduced within the statistical uncertainties
between the $b$-tagged and non-$b$-tagged samples.
In the $Z(\rightarrow ee)$+jets channel the multijet 
enriched control region is defined by following the full signal event selection with the exception of 
electron candidate impact parameter requirements, and requiring that one reconstructed electron candidate  
fails the `medium' quality requirement. As requirements based on the shower shape and associated 
ID track are applied to both electrons at trigger-level in the default trigger, events for the control region
are selected with a trigger which requires only a single electron with \et~$>20$~\gev. 
This trigger was only available for about one third of the full 2011 data-taking period (1.7~\ifb\ in total).
 In the $Z(\rightarrow\mu\mu)$+jets channel the multijet-enriched 
control region is defined by following the full signal event selection with the exception of muon candidate 
impact parameter requirements, and inverting the isolation selection for both reconstructed muon candidates. 
In both channels, contributions from non-multijet sources in the control regions are taken from simulation, and 
subtracted from the data. The remaining distributions are used as shape templates for the dilepton invariant 
mass distribution of the multijet background. 

Fits to $m_{\ell\ell}$ are then made after applying the full signal event selection, fixing the multijet
shapes to those measured in the control regions. For \onetag\ events the multijet contribution 
is determined to be 0.1$\pm$0.1\% in the electron channel and 0.02$\pm$0.07\% in the muon channel. 
The control regions are investigated as a function of all variables used to define the differential cross-sections
measured here, and no significant 
variation in the multijet fraction is found; therefore, 
the measured multijet fractions are assumed to be constant in all differential analysis bins. 
For \twotag\ events the multijet contributions are fitted to be zero, with an uncertainty of
approximately 0.5\%. This uncertainty is taken as a systematic uncertainty to account for a 
possible residual multijet contribution, as discussed in Section~\ref{sec:Systematics}.

\section{Extraction of detector-level signal yields}
\label{sec:Signal}

The extractions of the integrated and differential detector-level signal yields
for both the \onetag\ and \twotag\ selections are performed using maximum-likelihood 
fits to data based on flavour-sensitive distributions. 
The distribution used is constructed from the output of a 
neural network algorithm, JFComb, which is one of the inputs to the MV1 $b$-tagging algorithm 
described in Section~\ref{sec:Selection}.
JFComb itself combines the information from two further algorithms, 
one of which aims to identify weak $b\rightarrow c$ cascade topologies
using secondary vertices and displaced tracks reconstructed within a jet~\cite{jetfitter}, 
and the other which calculates a likelihood based on the impact parameter 
significance with respect to the PV of pre-selected tracks within \dR~$=0.4$ 
of the jet centroid~\cite{btag, btag2}. 
The JFComb algorithm has three outputs in the range [0,1]: $pb$, $pc$, and $pu$, 
corresponding to the probability that a given jet is a $b$-jet, $c$-jet or
 light-jet, respectively. Combinations of these variables, namely 
CombNNc~$=\ln(pb/pc)$ and CombNN~$=\ln(pb/pu)$ provide good separation between 
jet flavours as shown in Figure~\ref{fig:CombNNshapes} for all jets in the $Z$+jets MC 
simulation after the \onetag\ event selection. 

\begin{figure}[h]
\begin{center}
\subfigure[ ]{
  \includegraphics[width= 0.47\textwidth]{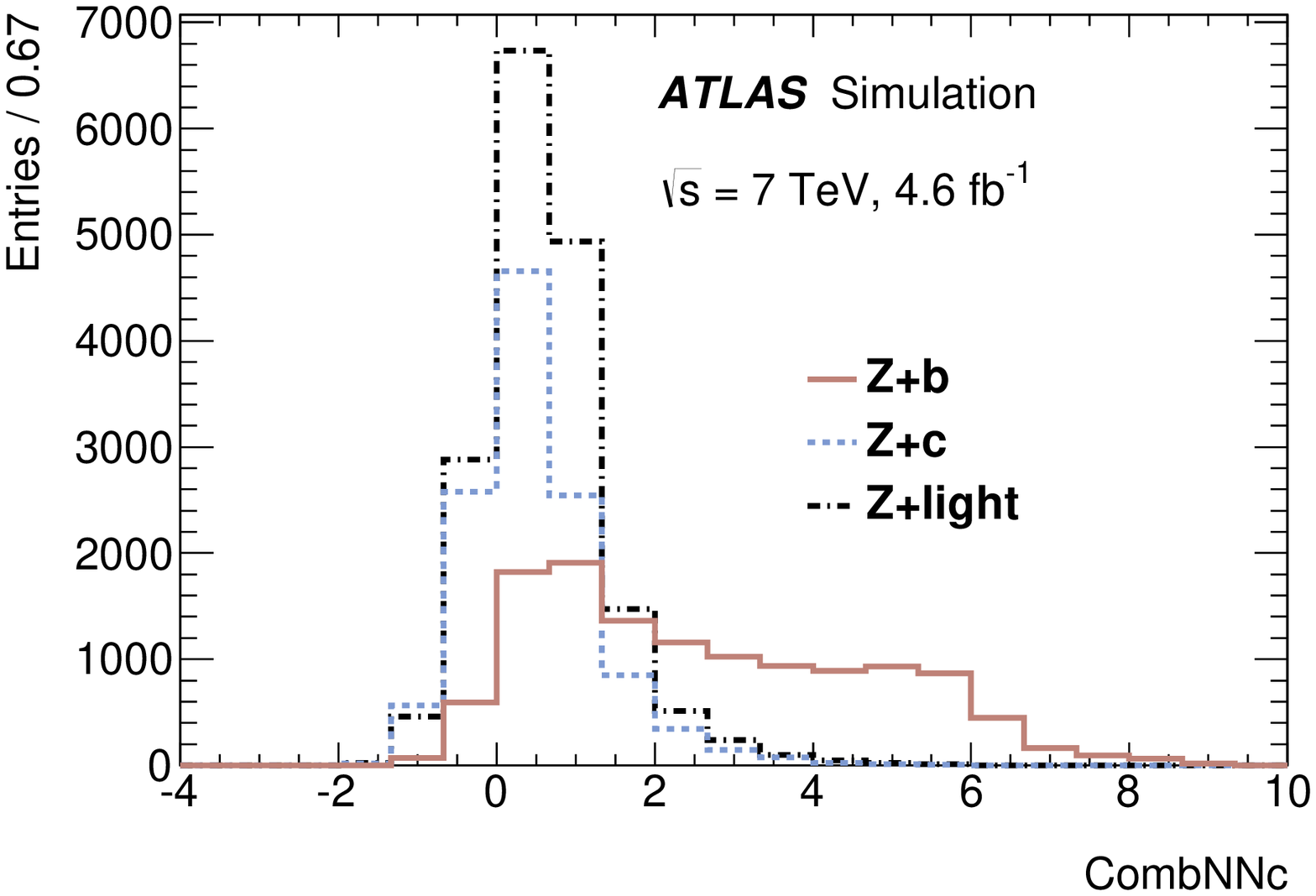}
  \label{fig:Zb_combnnc}
}
\subfigure[ ]{
  \includegraphics[width=0.47\textwidth]{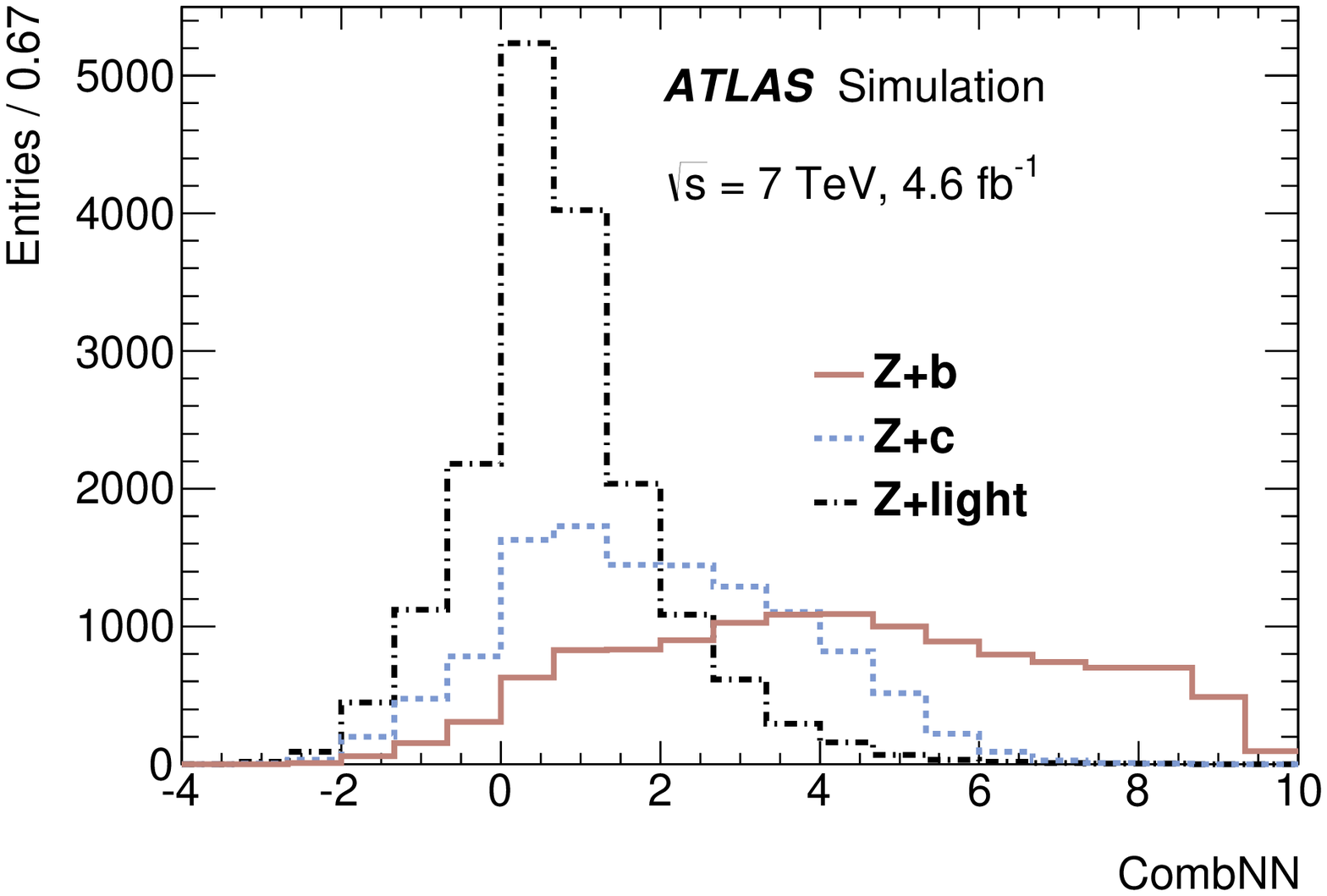}
  \label{fig:Zb_combnn}
}
\end{center}
\caption{Distributions of (a) CombNNc and (b) CombNN for different jet flavours in simulated $Z$+jets 
events for all selected tagged jets, in events with at least one tagged jet. The $Z\rightarrow ee$ 
and $Z\rightarrow\mu\mu$ channels are combined and simulated data are normalised such that the 
predicted number of jets in 4.6~\ifb\ are shown.
\label{fig:CombNNshapes}}
\end{figure}

In the \onetag\ event selection, fits are made to the CombNNc distribution
 as it is found to give the best
 statistical separation between $b$-jets and non-$b$-jets. Templates are
derived from MC simulation for all non-multijet contributions. For the multijet background, templates are 
derived from the respective control regions in each lepton channel
after reintroducing the $b$-tagging requirement as in the baseline selection. As shown in Figure~\ref{fig:Zb_combnnc}, 
the $c$-jet and light-jet CombNNc shapes are very similar.
They are therefore combined into a single non-$b$-jet template before the fit, using the predicted $c$-to-light jet ratio from simulation. 
Fits to data allow the $b$- and non-$b$-jet
$Z$+jets yields to float, while backgrounds from sources other than $Z$+jets are combined into a single template whose normalisation is determined from the sum of their predicted contributions and fixed in the fit. 
Where a per $b$-jet yield is measured, all tagged jets are
used in the fit; where a per-event yield is measured, only the highest \pt\ tagged jet in an event is used in the fit.
The electron and muon channel templates in data are combined before the fit to maximise the statistical precision. 
For measurements of differential cross-sections, these fits are performed independently in each bin, and Figure~\ref{fig:fit_example_Zb} shows an example fit to the CombNNc distribution in one differential bin, which is typical of the results obtained. 
Table~\ref{tab:Signal:detres} summarises all signal and background contributions compared to data for the integrated \onetag\ selections at detector-level after the jet-flavour fits. Also shown in Table~\ref{tab:Signal:detres} are the \alpgen+\herwig+\jimmy\ \onetag\ predictions, where it can be seen that they significantly underestimate the fitted $b$-jet yields.

\begin{figure}[t]
\begin{center}
\subfigure[ ]{
  \includegraphics[width= 0.47\textwidth]{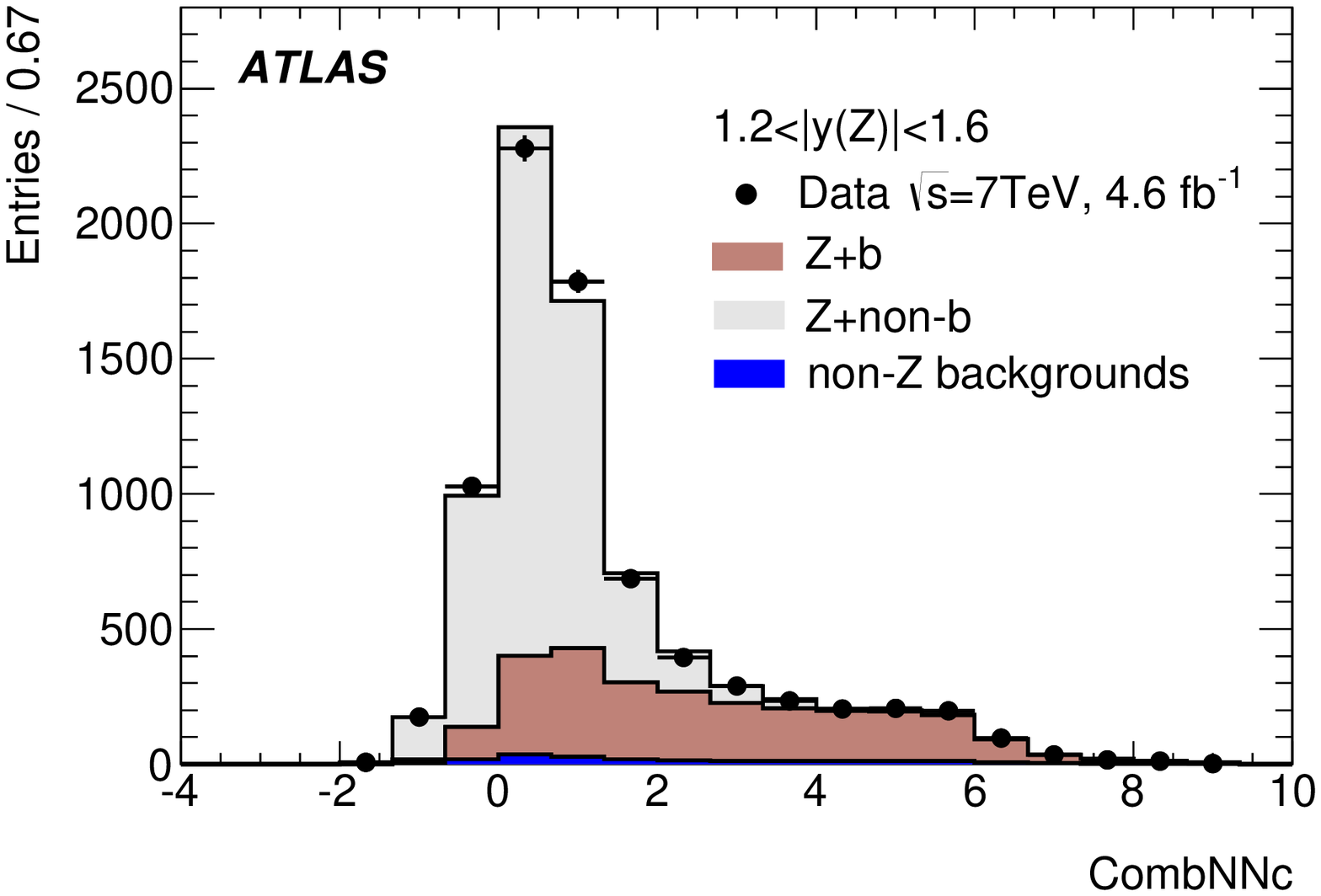}
  \label{fig:fit_example_Zb}
}
\subfigure[ ]{
  \includegraphics[width=0.47\textwidth]{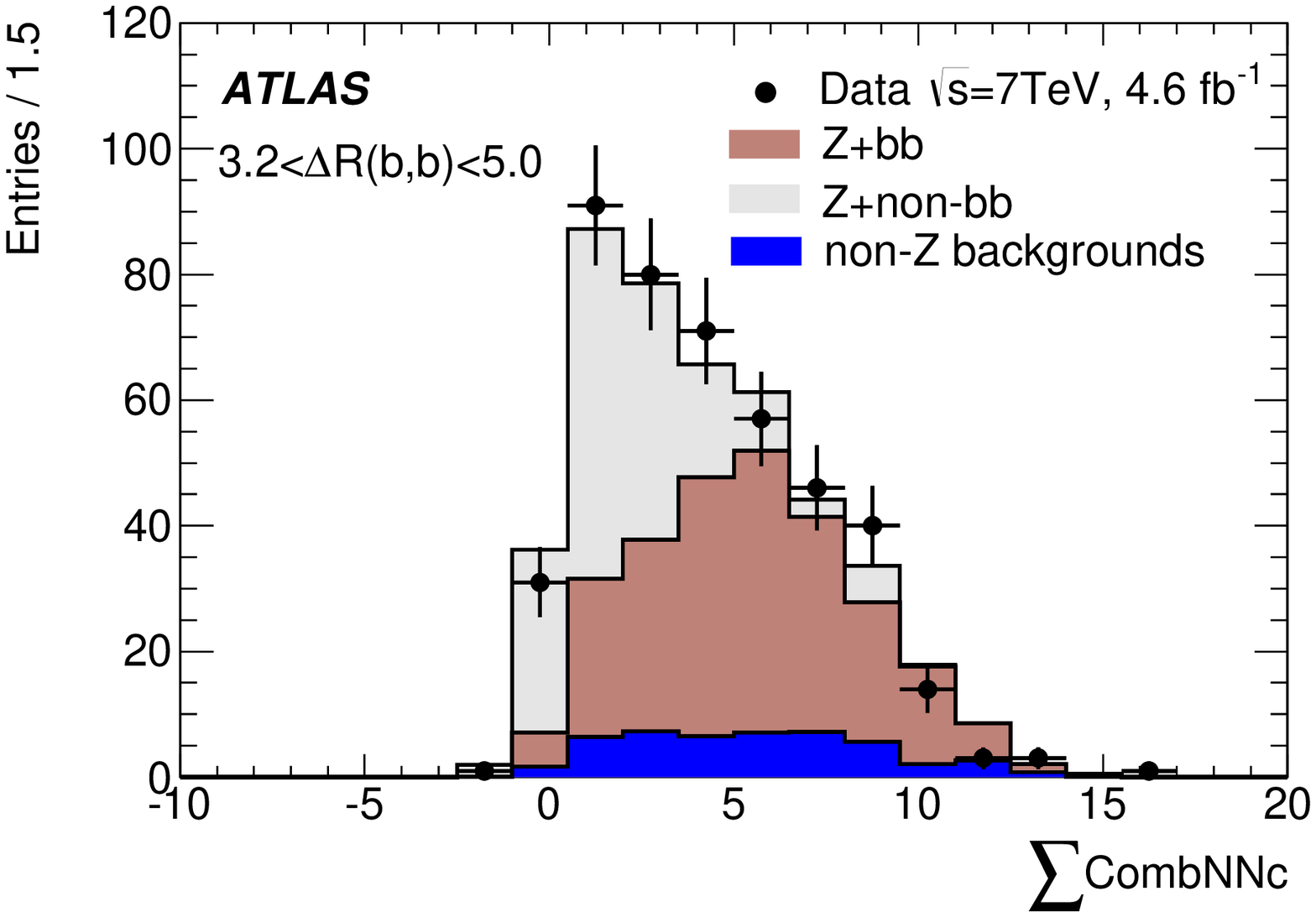}
  \label{fig:fit_example_Zbb}
}
\end{center}
\caption{Example fits to the distribution of (a) CombNNc at jet-level for 1-tag events with $1.2<|y(Z)|<1.6$, and (b) $\sum(\mathrm{CombNNc})$ at event-level for 2-tag events with $3.2<\Delta R(b,b)<5.0$.
\label{fig:fit_example}}
\end{figure}

In the \twotag\ event selection fits are made to $\sum(\mathrm{CombNNc})$, where 
the sum is over the two highest \pt\ tagged jets in an event.
There are six possible flavour combinations of $b$-jets, $c$-jets, and light-jets 
in the $Z$+jets MC simulation. 
The highest statistical precision on the signal $bb$-yield is obtained when 
the other five flavour combinations are combined into a single non-$bb$ template.
However, the shapes of the non-$bb$ templates are not degenerate, as the 
presence of a single $b$-jet in the $b$+light or $b+c$ cases results in a higher 
value of $\sum(\mathrm{CombNNc})$\ compared to light+light, light+$c$ and $c+c$ cases. 
Therefore, the overall number of these single-$b$ events is important in determining the
shape of the non-$bb$ template.
As discussed above, and can be seen in Table~\ref{tab:Signal:detres}, the 
\alpgen+\herwig+\jimmy\ simulation is observed to  underestimate the $b$-jet 
yield in data, and it follows that the number of $b$+light and $b+c$ events 
cannot be taken directly from the simulation when 
forming the non-$bb$ template, but must be measured. To determine the 
appropriate scaling for the templates containing a single $b$-jet, 
a fit is performed to CombNN in an alternative sample containing a reconstructed 
$Z$ boson with at least two jets, of which exactly one is tagged.
The $b$-jet, $c$-jet and light-jet $Z$+jets yields are allowed to float in the 
fit, while all non-$Z$+jets backgrounds are combined into a single template whose 
normalisation is determined from the sum of their predicted contributions and fixed 
in the fit; the multijet yields and shapes in this sample are extracted in a fashion 
analogous to that used for the \onetag\ events. 
Even after scaling up the total $Z$ cross-section to the NNLO prediction as described in Section~\ref{sec:Simulation},
the predicted $b$-jet yield must be increased by an additional factor of 1.35$\pm$0.03 
to match the fitted data yield, where the quoted uncertainty is the statistical component of the fit to data. 
Scale factors for $c$-jet and light-jet yields are found to 
be consistent with unity. Based on this result, templates containing one $b$-jet 
are weighted by a factor 1.35 compared to the predicted cross-section, 
while templates with no $b$-jets are included using the default predicted cross-section. 
This factor of 1.35 is taken as constant across all distributions as, 
normalisation aside, the default simulation is found to give a good description 
of the kinematics of the single $b$-jet sample. This scale factor is 
slightly different from the \Zb\ scale factors in Table~\ref{tab:Signal:detres},
due to the different jet requirements. A systematic uncertainty on the scale factor is 
obtained by varying these requirements, as described in Section~\ref{sec:Systematics}.

Signal fits to data float the $Z$+jets $bb$ and non-$bb$ yields 
while combining all other backgrounds into a single template whose normalisation is 
determined from the sum of their predicted contributions and fixed in the fit. 
As with \onetag\ events, the electron and  muon channels are combined 
before fitting to data to maximise the statistical precision. 
Figure~\ref{fig:fit_example_Zbb} shows an example fit of $\sum(\mathrm{CombNNc})$ 
in one differential bin, and Table~\ref{tab:Signal:detres} summarises all signal 
and background contributions compared to data for the integrated \twotag\ 
selection at detector-level after the jet-flavour fit.

All fits are checked with ensemble tests using the simulated samples, including checks for any bias
 in the fit results compared to the true number of $b$-jets in the simulation. Negligible
biases in the fit responses are observed.

\begin{table}[h]
\begin{center}
\begin{tabular}{|c|c|cc|cc|}
  \hline
  Analysis      & Data  & \multicolumn{2}{c|}{Fitted Components}    & \multicolumn{2}{c|}{Fixed Components} \\
  selection     & yield & $Z+b(b)$-jets [ALPGEN+HJ] & $Z$+(other)jets & \ttbar & other  \\ 
  \hline
  \sigZbjet     & 49701 & 18010$\pm$210 [12470] & 29780$\pm$230     & 1330   & 590  \\
  \sigZbjetstar & 41243 & 15640$\pm$190 [10460] & 23840$\pm$210     & 1230   & 540  \\
  \sigZb        & 47138 & 16610$\pm$200 [11410] & 29090$\pm$220     & 930    & 520  \\
  \sigZbb       & 2494  & 1170$\pm$60   [950]   & 860$\pm$50        & 395    & 60   \\
  \hline
\end{tabular}
\caption{Detector-level yields for each analysis selection. Statistical uncertainties from the fits 
to data are shown for the signal and $Z+$jets backgrounds. The \ttbar\ and other (diboson, 
single top quark and multijet) background normalisations are also shown. 
The signal yields predicted by \alpgen+\herwig+\jimmy\ (ALPGEN+HJ) are shown in square brackets for reference.}
\label{tab:Signal:detres}
\end{center}
\end{table}

\section{Correction to particle-level}
\label{sec:Unfolding}
Signal yields fitted at detector-level are corrected for reconstruction efficiencies and detector resolution effects using simulation. 
This unfolding procedure determines fiducial particle-level yields in data, which when divided by the measured integrated luminosity determine cross-sections.
Particle-level objects are selected with requirements chosen to be close to the corresponding requirements for reconstructed signal candidate detector-level objects, in order to minimise unfolding corrections.
Final state electrons and muons are `dressed', such that the four-momentum of collinear photons within \dR~$=0.1$ from those leptons are added to their four-momentum. 
These dressed leptons are then required to have $\pt>20$~\gev\ and $|\eta|<2.5$. The two leptons with highest \pt, same flavour and opposite charge are used to reconstruct the $Z$ boson, 
with the invariant mass of the pair required to lie in the range $76<m_{\ell\ell}<106$~\gev.
Jets of particles, excluding leptons used to reconstruct the $Z$ boson and any photons used in dressing them, 
but including leptons and neutrinos from heavy flavour decays, are reconstructed with the anti-$k_{t}$ algorithm
 with radius parameter $R=0.4$. As with simulated reconstructed jets, particle-level jets are defined as $b$-jets if they lie within \dR~$=0.3$ from one or more weakly decaying $b$-hadrons with $\pt>5$~\gev. Selected jets are required to have $\pt>20$~\gev\ 
and $|y|<2.4$.
Jets within \dR~$=0.5$ of a lepton used to reconstruct the $Z$ boson are discarded.

The classification of simulated signal events is based on the presence of detector-level and particle-level objects, 
and matching criteria between the two are defined.
The matching criteria require that detector-level and particle-level event selections are passed and that each detector-level $b$-jet lies within \dR~$=0.4$ from a particle-level $b$-jet. For event-level (jet-level) differential measurements, matched events ($b$-jets)
are used to populate detector response matrices for the distribution in question. These matrices characterise the bin migrations between detector-level and particle-level quantities
and are used to unfold the fitted signal yields at detector-level into signal yields at particle-level. 

Before unfolding, a multiplicative matching correction derived from simulation is 
applied to the fitted signal yields, to account for cases where the detector-level signal failed 
the matching criteria. This correction is 6--9\% for the integrated selections, 
although it becomes as large as 20\% in the lowest bin of $b$-jet \pt\ in the \onetag\ analysis 
due to migration from particle-level $b$-jets below the 20~\gev\ \pt\ threshold. In order to avoid bias in the differential cross-section measurement of $b$-jet \pt, detector-level $b$-jets are considered 
as matched if they are associated to particle-level $b$-jets with $\pt>10$~\gev. 
For other 
variables the migration outside of acceptance is found to introduce negligible bias and hence the particle-level $b$-jet selection
is only relaxed in the unfolding of $b$-jet \pt. For \twotag\ events, where simulation sample size for matched events is a 
limiting factor, the $b$-tagging efficiency correction is included as part of the matching 
correction. This allows all detector-level $b$-jets, tagged or otherwise, to be used in the 
response matrices.

In the \onetag\ analysis, corrected fitted yields and response matrices are used as input to an
iterative Bayesian technique \cite{RooUnfold} to extract the particle-level signal yields. 
Three further iterations on the initial response matrix are required to remove bias 
from previous iterations as determined from MC simulation ensemble tests of the statistical robustness of the unfolding procedure. 
The binning of differential distributions is chosen to always be significantly wider than the detector resolution
in that quantity, which is only a relevant factor for $b$-jet \pt. Related to this resolution effect, and to again 
mitigate the biases mentioned above, the response matrix  for $b$-jet \pt\ is also populated with 
particle jets with  $\pt>10$~\gev, and the portion of the resulting particle-level distribution below 20~\gev\ is removed.
In the \twotag\ differential distributions, fewer events are selected and binnings are chosen to optimise statistical precision
while maintaining as many bins as possible. This coarse binning results in little bin-to-bin migration,
and a negligible difference is observed between the result of the iterative procedure used for \onetag\ events and that obtained by simply applying fiducial matching and efficiency
corrections\footnote{The efficiency correction refers to the multiplicative factor obtained from the ratio of the total number of fiducial particle-level events to the number which is also reconstructed at the detector-level and matched appropriately.} individually for each bin. As a consequence, the latter, more straightforward, technique is used to extract differential yields
in \twotag\ events.

Since the electron and muon $Z$ boson decay channels are combined to increase the precision of the signal fits to data,
the corrections and response matrices described above must 
unfold both channels simultaneously to obtain combined particle-level yields. To 
validate this approach, an identical analysis of each individual lepton channel is performed. 
Their sum after unfolding is checked for consistency with the default combined unfolded result 
and excellent agreement is observed in both the \onetag\ and \twotag\ cases.
Furthermore, the results obtained from the individual lepton channels agree reasonably well, 
being compatible within $1.7\sigma$ or less,
considering only the sum in quadrature of the statistical and uncorrelated systematic uncertainties.

\section{Systematic uncertainties}
\label{sec:Systematics}
Several sources of systematic uncertainty are considered. These can
impact either the fit, through modification of template shapes and background
 normalisations; the unfolding, through modification of correction factors
and response matrix; or both the fit and unfolding in a correlated manner. 
Each independent source of uncertainty is varied successively up, and then down by 
one standard deviation, and in each case the full analysis chain is repeated.
The relative change in the result with respect to the default analysis is 
then assigned as the up or down uncertainty due to that source. The following sources
are considered and the resulting uncertainties on the measured \onetag\ and \twotag\ cross-sections are summarised in Table~\ref{tab:systematics}.

{\bf Tagging efficiency and mistag rates.} Calibration factors are applied to the jet $b$-tagging 
efficiency in simulation to match that measured in data for each flavour. These have associated systematic
uncertainties as a function of jet \pt\ (and $\eta$ for light-jets). For $b$-jets, the uncertainties 
derived from calibration analyses are divided into 10 sub-components corresponding to the eigenvectors 
which diagonalise the associated covariance matrix; each sub-component is then varied independently by 
$\pm1\sigma$ and the systematic uncertainties from each are added in quadrature. Typically two of the sub-components 
dominate the uncertainty, with one at around 4.5\% at low $b$-jet \pt, and the other rising to 
around 13\% at high $b$-jet \pt. Across other distributions, both remain between 2-3\%.
For $c$-jets and light-jets, the mistag correction factors from each respective calibration analysis 
are varied by $\pm1\sigma$ and propagated through the analysis chain to obtain the corresponding systematic uncertainties.
These contribute significantly smaller uncertainties, peaking at around 1\% at high $b$-jet rapidity and low \pt.
All uncertainties related to $b$-jets approximately double in size when requiring two tagged jets for the \Zbb\ 
distributions.

{\bf Jet energy scale.} Systematic uncertainties on the difference between the jet energy scale (JES) in 
data and simulation are derived using a variety of studies based on in situ measurements and simulation~\cite{jes2011}.
 These uncertainties are decomposed into 16 independent components, 
including those arising from the influence of close-by low-energy jets, the correction for pile-up 
activity and differences in detector response to light-quark jets, gluon jets and heavy flavour jets. Each component 
is propagated through the analysis chain independently by simultaneously varying the signal and background
simulation jet response by $\pm1\sigma$. 
The impact of the total JES uncertainty on the final cross-sections is typically around 2-5\%, rising with \pt\ and rapidity, with uncertainty on the $b$-jet response uncertainty being an important contribution.

{\bf Jet energy resolution.} Jet energy resolution (JER) is studied 
in dijet data and compared to simulation \cite{JERsys}. Simulated signal and background samples are then 
modified by applying a Gaussian smearing of the resolution function according to the maximum degradation 
allowed by the JER measurement from data to evaluate the associated systematic variation. This is taken as
a symmetrised uncertainty on the measured results, and is typically less than 1\%.

{\bf \emph{b}-jet template shapes.} The uncertainty on the shape of $b$-jet templates used in fits to data 
is a dominant contribution to the overall systematic uncertainty for this analysis. The shape is cross-checked 
in a \ttbar-enriched control region which requires a single well identified and isolated lepton in association 
with at least four reconstructed jets passing the same kinematic cuts as signal jets, of which exactly two are 
tagged with the MV1 algorithm described in Section \ref{sec:Selection}.
This selects a sample of \ttbar\ events in which over 90\% of the tagged jets are expected to be true $b$-jets. 
Contributions from $W$+jets and single top quark events in this control region are estimated from simulation; 
contributions from other electroweak processes and multijet backgrounds are found to be negligible. 
A residual underestimate of the total number of events predicted by simulation is found to be less than 
10\%, and is corrected for by scaling up the \ttbar\ contribution to match the data normalisation. 
Figure~\ref{fig:TopControl} shows the CombNNc distributions for different jet flavours, and the ratio of data to default 
simulation for all tagged jets in this 
control region; the corrections of the \herwig\ $b$-hadron decays to the \evtgen\ prediction described in 
Section~\ref{sec:Simulation} are applied. It can be seen that the simulation provides a reasonable description of 
the data; the residual differences of up to 5\% are used to derive a $b$-jet template reweighting function shown by the 
dashed line in Figure~\ref{fig:TopControl}. Investigation of the control-region data in
bins of tagged jet \pt\ and rapidity finds that the deviations between data and simulation have no strong
 dependence on tagged jet kinematics; this is despite the CombNNc distribution shape itself having a strong 
dependence on tagged jet \pt. The reweighting function is used to directly reweight CombNNc $b$-jet distributions 
in the signal $Z$+jets simulation and the fits to data are repeated. The relative differences with respect to the default 
results are typically less than 5\%, and this difference is taken as a systematic uncertainty due to $b$-jet template shape, which is then symmetrised around 
the nominal value to give an up- and down-uncertainty. It is possible that the $b$-jet template reweighting function derived reflects some mismodelling of the non-$b$-jets component in Figure \ref{fig:TopControl}. To this end the systematic uncertainty is also evaluated by only reweighting $b$-jet templates for values of CombNNc larger than 2.0. The result is a smaller overall uncertainty; however, as the fraction of $b$-jets is still larger than the fraction of non-$b$-jets for CombNNc less than 2.0 in Figure \ref{fig:TopControl}, the more conservative value obtained from reweighting the full $b$-jet template is taken as the uncertainty. As a further cross-check, the fits are repeated 
using $b$-jet templates obtained from the \sherpa\ $Z$+jets sample; deviations observed are all within the uncertainties 
already derived from the \ttbar-enriched control-region method, so no further uncertainty is assigned.

\begin{figure}[h]
 \centering
    \includegraphics[width=0.7\columnwidth]{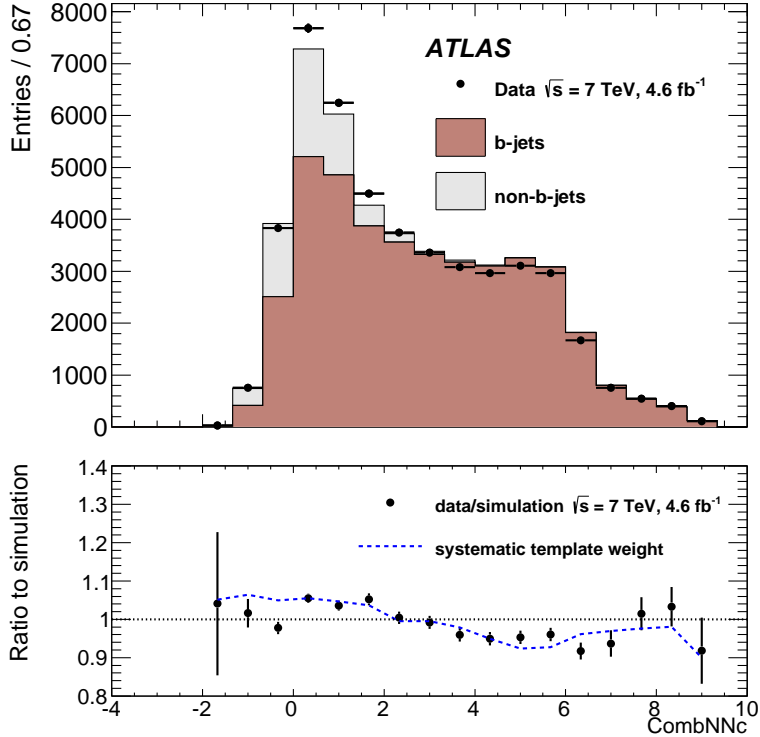}
    \caption{
The tagged-jet CombNNc distribution in the \ttbar\ enriched control region described in the text, 
with the simulation split by jet flavour (top), and the ratio of data to default simulation (filled circles, bottom).
The dashed line shows the $b$-jet template reweighting function derived from the difference between 
data and simulation in this control region.
  \label{fig:TopControl}}
\end{figure}

{\bf Non-\emph{b}-jet template shapes.} Mismodelling of template shapes derived from $Z$+jets 
simulation for $c$-jets and light-jets can also cause a systematic shift in the results of fits 
to data. The corresponding uncertainties are estimated by substituting 
the default templates with templates derived from the \sherpa\ $Z$+jets simulation, which uses a 
different parton shower and hadronisation model. The difference between 
the default fit response and the response obtained with the alternative templates is taken 
as the systematic uncertainty, which is typically less than 1\%.

Further tests are made by repeating the entire analysis using an MV1 operating point which rejects significantly more 
$c$- and light-jet background (but with a lower signal efficiency), reducing the sensitivity to any potential 
mismodelling of these templates. The results are entirely consistent with those obtained using the default value, so no 
further uncertainty is assigned.

Finally, the template shapes may be influenced by a mismodelling of the light- and $c$-jet kinematics,
and by the light-/$c$-jet ratio in the simulation when building the non-$b$ template. The data are 
fitted using CombNN rather than CombNNc (defined in Section \ref{sec:Signal}),
which provides a better discrimination between light- and $c$-jets, and all three templates (light-, $c$- and $b$-jet)
are allowed to float. Across all distributions, the fitted light- and $c$-jet normalisations are consistent with the prediction 
of the default simulation within the statistical uncertainties (typically 2-4\%), indicating the kinematic modelling of these contributions, and their ratio, is correct. Therefore
no further systematic uncertainty is assigned.

{\bf Template scale factor.} The $b$+$c$ and $b$+light jet templates in \twotag\ events are scaled up by a 
factor of 1.35, as described in Section~\ref{sec:Signal}, based on fits to data with two or more jets, of 
which exactly one is tagged. A fit to integrated \onetag\ data yields a factor of 1.48; the default scale 
factor of 1.35 is varied up and down by 0.13 to cover this difference, resulting in a change in the final cross-sections
of around 2\%, which is assigned as a symmetric up and down systematic uncertainty. 
The $c$- and light-jet fractions in these templates are also independently varied up and down by 0.13, 
significantly larger than the fit uncertainties and differences in the two control regions used to derive 
the $b$-jet scale factor, but chosen to based on the $b$-jet result to provide a conservative bound on 
mismodelling of the $c$-jet fraction. This results in a further systematic uncertainty of around 1\%.

{\bf Multiple parton interactions.} Associated $Z$+$b$-jets production from MPI where the $Z$ and $b$-jets are produced in separate
hard scatters within a single $pp$ interaction (double parton interactions) is included in 
the analysis signal definition. Fits to data and unfolding to particle-level use 
the MPI fraction predicted by \jimmy\ in $Z$+jets simulation to determine its relative contribution to the signal processes. The contribution is largest 
at lower $b$-jet \pt; any misestimate of this fraction can alter the CombNNc shapes, 
which are \pt -dependent, and can also alter the efficiency correction and the bin-by-bin migration
 in \pt -dependent variables. 
The default double parton interaction fractions as a function of $b$-jet \pt\ and rapidity are cross-checked by 
combining ATLAS measurements of the $Z$ boson production cross-section \cite{InclZ2010}, 
the differential inclusive $b$-jet cross-section \cite{Diffb2010} and $pp$ effective cross-section \cite{EffxSec2010} using
the phenomenological model described in reference \cite{EffxSec2010}. 
The prediction from \jimmy\ is found to be consistent with this data-based 
cross-check to within 50\%, hence the predicted fraction is varied by this amount to determine an associated
systematic uncertainty. The uncertainty is typically around 2\% on the measured cross-sections.

{\bf Gluon splitting.} The dominant mechanism to produce two $b$-hadrons in one jet is the $g\rightarrow b\overline{b}$ 
process. An inaccurate estimate of the rate of two $b$-hadron decay vertices within \dR~$=0.4$ from 
the jet centroid can affect the accuracy of the CombNNc template shapes, by impacting distributions which are 
inputs to the NN. Furthermore, as gluon 
splitting becomes more important for high \pt\ jets, a mismodelling of its rate can impact the efficiency
 correction and bin migrations in variables correlated with $b$-jet \pt. No well defined data control region 
has been identified to constrain this process; therefore
the sample of simulated events with reconstructed and particle-level jets matched to two $b$-hadrons is first
enhanced by a factor of 2, then completely removed, with the full analysis being repeated in both cases. 
This variation is larger than the difference observed between predictions from the default 
signal \alpgen+\herwig+\jimmy\ and \sherpa\ $Z$+jets simulations and is therefore considered to give a 
conservative upper limit on the magnitude of this uncertainty, and is found to be less than 2\%.

{\bf Background normalisation.} The contributions of \ttbar, single top quark and diboson backgrounds are taken from theoretical predictions. To account for theoretical uncertainties in these predictions 
the normalisation of each component is varied independently by $\pm$10\%, which covers both cross-section 
and acceptance uncertainties. For \onetag\ events the multijet background is varied within its fitted uncertainty.
For \twotag\ events, fits for the multijet backgrounds yielded a normalisation close to zero, and the uncertainty from those fits is taken 
as an upper bound for possible multijet contamination, translating into an uncertainty of 0.5\% on the $bb$ yield.

{\bf Background modelling.} A cross-check of \ttbar-background modelling is made by substituting the default MC 
simulation with an alternative sample simulated with \powheg\ and repeating the data fits. For \onetag\ events 
no significant difference is found, either inclusively or differentially. In \twotag\ events a systematic 
deviation in excess of the existing template-shape uncertainty described above is observed. 
This difference is approximately 3\%, which is taken as an additional systematic uncertainty due to \ttbar\ modelling in the \twotag\ sample.

{\bf Signal modelling.} The corrections to particle-level cross-sections may include some residual dependence 
 on the modelling of the kinematics in the simulation. To test for this, the particle-level $b$-jet \pt\ distribution
in simulation is reweighted to the measured differential cross-section, and the full analysis repeated. A negligible effect is 
found. As the main kinematic distributions are generally well modelled by the simulation, no further 
uncertainties are assigned.

{\bf Simulation sample size.} The impact of the finite simulation sample sizes in both the fit template shapes 
and unfolding procedure are evaluated through ensemble tests, repeating the analysis and randomly fluctuating
bin entries of a given distribution in the simulation within their statistical uncertainty. The spread determined 
from these ensemble tests is around 1\%, which is assigned as the systematic uncertainty.

{\bf Lepton efficiency, energy scale and resolution.} 
The trigger and reconstruction efficiency, energy scale, and resolution of both reconstructed electron and 
muon candidates have been measured in $Z\rightarrow\ell\ell$ events and used to correct the simulation as described in Section~\ref{sec:Selection}. The uncertainties associated with the measurement of these quantities are 
propagated through the full analysis chain resulting in an uncertainty of around 1\% on the final cross-sections.

{\bf Missing transverse momentum.} The calculation of \met\ in each event is repeated for every systematic 
variation of reconstructed jet and lepton candidates as described above. An additional uncertainty arises from 
possible differences in data and simulation between the component of \met\ from topological clusters not 
associated to reconstructed physics objects \cite{MET2011}. This additional component is propagated through 
the analysis as an independent uncertainty, and is typically well below 1\%.

{\bf Luminosity.} The luminosity scale is determined from a single calibration run taken in May 2011. The associated 
uncertainty is derived from the calibration analysis itself and from the study of its stability across the 2011 
data taking period. A total uncertainty of 1.8\% is assigned to the luminosity \cite{lumipaper}.

\begin{table}[h]
\begin{center}
\begin{tabular}{|c|cc|}
  \hline
  Source of  &         &         \\
   uncertainty&  \sigZb\ [\%]  & \sigZbb\ [\%] \\ 
  \hline
  $b$-jet tagging efficiency & 3.4 & 9.8 \\
  $c$-jet mistag rate & 0.2 & 2.3  \\
  light-jet mistag rate & 0.4 & 0.6  \\
  JES & 2.9 & 4.7  \\
  JER & 0.3 & 0.7  \\
  $b$-jet template shape & 4.8 & 4.8  \\
  $c$-jet template shape & 0.2 & 0.6  \\
  light-jet template shape & 0.9 & 0.9  \\
  $b$-jet template scale factor & N/A & 2.3 \\
  MPI & 2.5 & 0.8 \\
  gluon splitting & 1.2 & 1.5 \\
  background normalisation & 1.1 & 3.6 \\
  \ttbar\ modelling & 0.0 & 2.9\\
  MC sample size & 1.0 & 1.4 \\
  lepton efficiency, scale and resolution &1.2 & 1.2\\
  \met & 0.1 & 0.6 \\
  luminosity & 1.8 & 1.8 \\
  \hline
  total & 7.7 & 14.0\\
  \hline
\end{tabular}
\caption{Summary of the systematic uncertainties determined for the cross-section measurements of the \Zb\ and \Zbb\ final states.}
\label{tab:systematics}
\end{center}
\end{table}

\section{Theoretical predictions}
\label{sec:Theory}

Several theoretical predictions are compared to the measurements. Fixed-order pQCD parton-level predictions at 
NLO in the 5FNS are obtained from \mcfm\ \cite{MCFM} for both the \Zb\ and \Zbb\ final states. The calculation of \Zb\ is made up of several 
sub-processes~\cite{MCFMZb, MCFMZbj} at $\mathcal{O}(\alphas^{2})$, and the $b$-quark mass is ignored except in processes where one $b$-quark 
falls outside the acceptance or two $b$-quarks are merged in a single jet.
For \Zbb\ production, the \mcfm\ calculation uses a single process with both $b$-quarks in acceptance at $\mathcal{O}(\alphas^{3})$ and the $b$-quark mass is ignored throughout. In all cases, the renormalisation and factorisation 
scales are set to $\sqrt{m(Z)^2 + p_T(Z)^2}$, and varied up and down independently by a factor of two to assess 
the dependence on this scale choice. 
The \mcfm\ predictions are performed using the CT10~\cite{ct10}, NNPDF2.3~\cite{nnpdf} and MSTW2008~\cite{mstw08} PDF sets. The uncertainties associated with the PDF fits to experimental data are propagated appropriately for each PDF set. 
The dependence on the choice of $\alphas(m(Z))$ is assessed by using PDF sets with $\alphas(m(Z))$ shifted up and down by the
 68\% confidence level interval around the default value used in the PDF.
For MSTW2008, fits using different $b$-quark masses are also available. 
The prediction from \mcfm\ is at the parton-level, so must be corrected for the effects of QED final-state radiation (FSR), hadronisation, underlying event and MPI. 
The correction for QED FSR is obtained using \photos, interfaced to the \alpgen+\herwig+\jimmy\ samples used in the data analysis, and evaluated by comparing the cross-sections obtained by applying the selection requirements to  leptons before, and after FSR.
The correction factors for hadronisation, underlying event and MPI are obtained for each differential cross-section from both \pythia\ and \sherpa, 
by taking the ratio of the predictions with these effects turned on and turned off. 
The versions used are \pythia\ 6.427, with the \cteqfl\ PDF set and the Perugia 2011 tune, and \sherpa\ 1.4.1, with the \ctten\ PDF set. 
Differences between the correction factors obtained in \pythia\ and \sherpa, which are typically at the 1\%-level, as well as the 50\% uncertainty on MPI described in Section~\ref{sec:Systematics}, are assigned as systematic uncertainties.

Full particle-level predictions with NLO matrix element calculations are also obtained using \amcnlo~\cite{aMCNLO}, in both the 4FNS and 5FNS. In the 4FNS, 
the \Zbb\ process is calculated at $\mathcal{O}(\alphas^{3})$, including the effects of the $b$-quark mass, and interfaced to the MSTW2008NLO\_nf4 PDF set~\cite{mstw08}. No kinematic cuts are applied to the $b$-jets in this calculation, therefore it is also used to derive a 4FNS prediction for the \Zb\ final state.
For the 5FNS prediction, the more inclusive $Z$+$\ge 1$-jet process is calculated at $\mathcal{O}(\alphas^{2})$ neglecting the $b$-quark mass and using the MSTW2008NLO PDF set. This is then used to derive a 5FNS prediction at $\mathcal{O}(\alphas^{2})$ for \Zb\ and \Zbb. The latter process is therefore LO only.
In both cases, \herwigpp\ is used to simulate the hadronisation, underlying event and MPI. 
Both predictions require a correction for a missing component of MPI, in which the $Z$ boson and 
$b$-quarks are produced in separate scatters within the $pp$ collision. 
This correction is estimated using the \alpgen+\herwig+\jimmy\ samples where the MPI contribution is included. Since the 4FNS and 5FNS 
use different matrix elements ($Z$+$bb$ and $Z$+jet respectively), a different correction factor is derived in each case. 
In both the 4FNS and 5FNS predictions from \amcnlo, the renormalisation  and factorisation scales are 
set dynamically to the same definition used for the \mcfm\ prediction.
Since variations of the scales are the dominant sources of theory uncertainty, they have been evaluated for all \amcnlo\ predictions using the same procedure as for \mcfm. 
The overall scale uncertainty is found to have a comparable size in the 4FNS and 5FNS predictions, and 
to be consistent with the scale uncertainty for \mcfm.  However, the uncertainty 
is fully dominated by variations of the renormalisation scale in the 4FNS case, while for the 5FNS  
renormalisation and factorisation scale variations produce shifts which are similar in magnitude and 
opposite in direction, giving a total uncertainty dominated by the cases where one is shifted up and the
other down (and vice versa). 
Uncertainties arising from the PDFs and the choice of $\alphas$ are obtained using \mcfm.

Predictions are also obtained from \sherpa\ and \alpgen+\herwig+\jimmy, which combine tree-level 
matrix elements for multiple jet emissions with a parton shower, hadronisation and underlying 
event package. \alpgen\ uses the 4FNS and has up to five partons in the matrix element, while \sherpa\ uses the 5FNS and has up to four partons in the matrix element.

\section{Results}
\label{sec:Results}

The cross-sections for \Zb\ and \Zbb\ are shown in Figure~\ref{fig:results_summary},
 and Table~\ref{tab:thxs}. The \mcfm\ predictions always agree with the data within the 
combined experimental and theoretical uncertainties. The prediction obtained 
with \ctten\ is lower, due primarily to the default choice of $\alphas(m(Z))$ in this PDF (0.118) compared 
to MSTW2008 and NNPDF2.3 (0.120 in each). The predictions do agree within the uncertainty on the 
choice of $\alphas(m(Z))$.
For \amcnlo, the 5FNS prediction describes \Zb\ well, while the 4FNS underestimates the measured cross-section. 
This situation is reversed for the \Zbb\ case, where the 4FNS provides a good description, while the 5FNS underestimates the cross-section.
However, as explained in Section~\ref{sec:Theory}, the 5FNS prediction from \amcnlo\ is only LO  for \Zbb, which may explain this underestimate.   
Considering only statistical uncertainties, both the 4FNS prediction from \alpgen+\herwig+\jimmy\ and the 5FNS prediction from \sherpa\ underestimate the data, with \alpgen+\herwig+\jimmy\ being consistently below \sherpa\ by around 30-40\%.

\begin{table}[h]
  \begin{center}
    \small
    \begin{tabular}{lcccc}
      \hline
      \hline
      & \sigZb [fb]   & \sigZbjet [fb] &  \sigZbjetstar [fb]  &   \sigZbb [fb] \\
      \hline
      Data & $4820\pm60^{+360}_{-380}$ & $ 5390\pm 60\pm480 $ & $ 4540 \pm 55\pm330 $ & $ 520\pm 20^{+74}_{-72} $ \\
      \hline
      MCFM$\otimes$MSTW2008 & $ 5230\pm 30^{+690}_{-710}$ & $5460 \pm 40^{+740}_{-740} $ & $    4331 \pm 30^{+400}_{-480} $ & $ 410\pm 10^{+60}_{-60} $ \\
      MCFM$\otimes$CT10     & $4850\pm30^{+580}_{-680}$ & $5070\pm 30 ^{+640}_{-710}$ & $4030\pm 30 ^{+350}_{-450}$  & $386 \pm 5^{+55}_{-50}$ \\
      MCFM$\otimes$NNPDF23 & $5420\pm 20^{+670}_{-710}$ &  $5660 \pm 30 ^{+720}_{-740}$ & $4490 \pm 30 ^{+380}_{-460}$ & $420 \pm 10 ^{+70}_{-50}$ \\

      \amcnlo\ 4FNS$\otimes$MSTW2008 & $ 3390\pm20^{+580}_{-480}  $ &  $ 3910\pm20^{+660}_{-560}  $  & $ 3290\pm20^{+580}_{-460}  $ &  $  485\pm7^{+80}_{-70}   $ \\
      \amcnlo\ 5FNS$\otimes$MSTW2008 & $4680\pm40^{+550}_{-580}$  & $5010\pm40^{+590}_{-620}$ & $4220\pm40^{+460}_{-510}$ &  $314\pm9^{+30}_{-30}$  \\
      \sherpa$\otimes$CT10      & $ 3770  \pm 10 $  & $ 4210  \pm 10 $  & $ 3640  \pm 10 $  & $ 422   \pm 2 $  \\
      \alpgen+HJ$\otimes$CTEQ6L1 & $2580 \pm 10$ &$2920 \pm 10$ &$2380 \pm 10$ &  $317 \pm 2$   \\
      \hline
    \end{tabular}
  \end{center}
  \caption{The measurement and theory predictions for the integrated cross-sections and the integrated inclusive $b$-jet cross-sections. The \mcfm\ results are corrected for MPI, non-perturbative QCD effects and QED radiation effects. The statistical uncertainty is quoted first in each case. The second uncertainity is either the total systematic uncertainty (data), the sum in quadrature of all theory uncertainties (\mcfm), or the scale uncertainty (\amcnlo).\label{tab:thxs}}
\end{table}

\begin{figure}[thp]
\begin{center}
\subfigure[ ]{
  \includegraphics[width= 0.8\columnwidth]{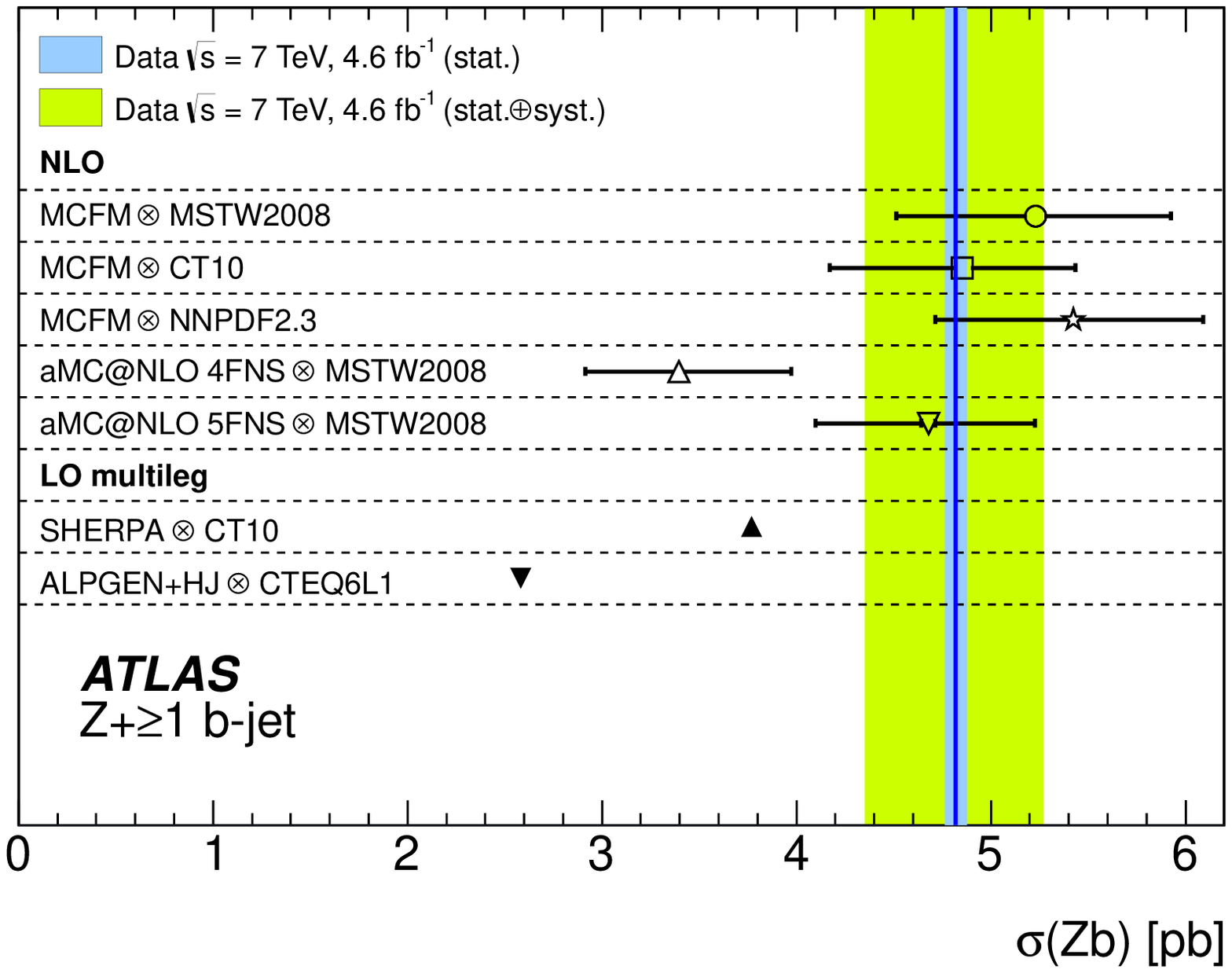}
  \label{fig:results_summary_Zb}
}
\subfigure[ ]{
  \includegraphics[width=0.8\columnwidth]{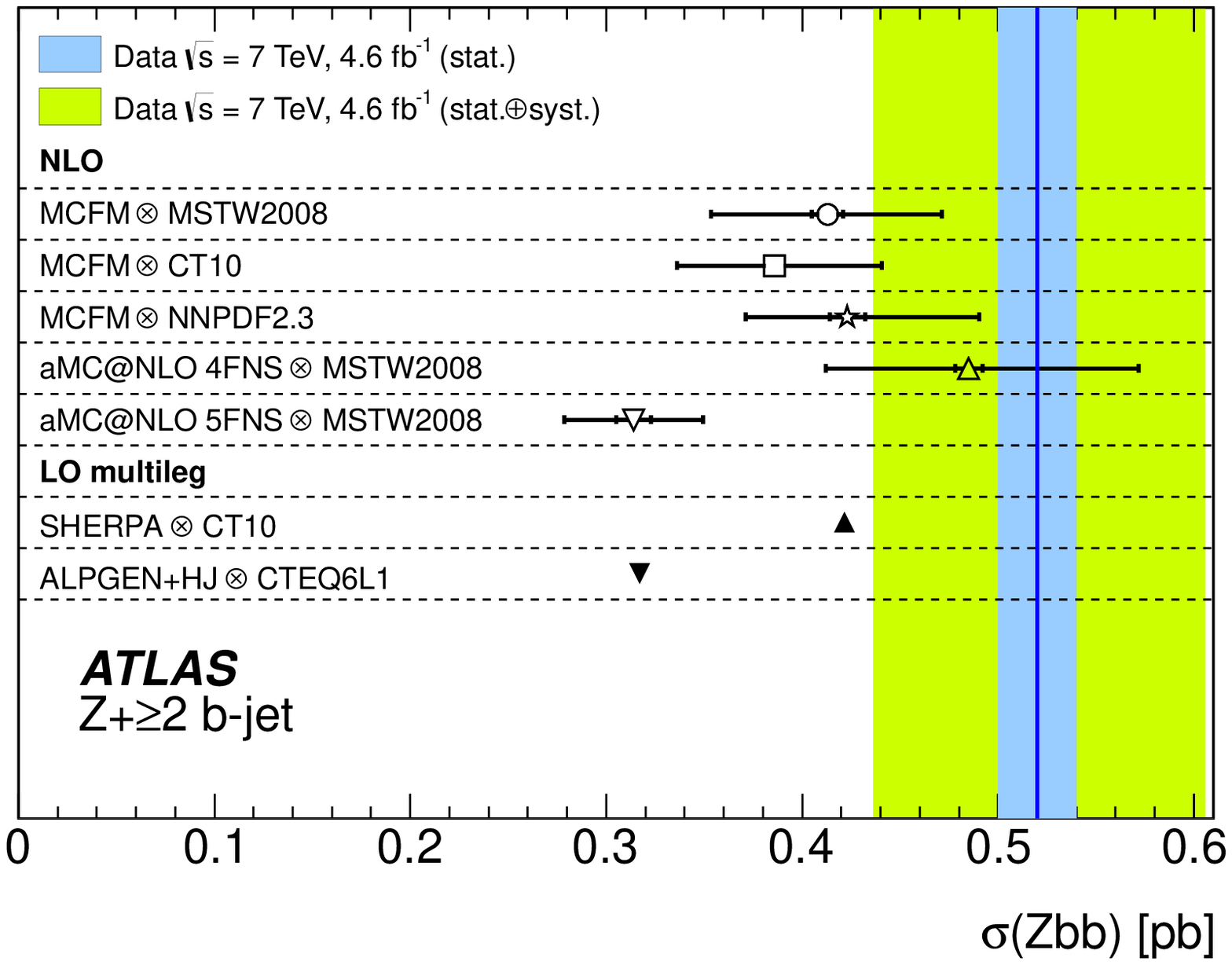}
  \label{fig:results_summary_Zbb}
}
\end{center}
\caption{Cross-sections for (a) \Zb, and (b) \Zbb. The measurement is shown as a vertical blue line with the 
inner blue shaded band showing the corresponding statistical uncertainty and the outer green shaded band showing the sum in 
quadrature of statistical and systematic uncertainties. Comparison is made to NLO predictions from \mcfm\ interfaced to different 
PDF sets and \amcnlo\ interfaced to the same PDF set in both the 4FNS and 5FNS. The statistical (inner bar) and total (outer bar) 
uncertainties are shown for these predictions, which are dominated by the theoretical scale uncertainty calculated as described in the text. 
Comparisons are also made to LO multi-legged  predictions from \alpgen+\herwig+\jimmy\ and \sherpa; in this case the uncertainty bars are 
statistical only, and smaller than the marker.
\label{fig:results_summary}}
\end{figure}

Figure~\ref{fig:results_bjet_pTy} shows \sigZbjet, as a function of the $b$-jet \pt\ and \absy. The theoretical predictions generally provide a good description of the shape of the data. The  4FNS prediction from \amcnlo\ underestimates the data most significantly at central rapidities.
Figure~\ref{fig:results_Z_pTy} shows \sigZb, as a function of the $Z$ boson \pt\ and \absy. In general, all theoretical predictions provide a reasonable description of the shape of the data within uncertainties, though there is evidence for disagreement at very high $Z$ boson \pt, and a slope in the ratio of the \mcfm\ prediction to data for the $Z$ boson rapidity.

\begin{figure}[hp]
\begin{center}
\subfigure[ ]{
  \includegraphics[width= 0.5\textwidth]{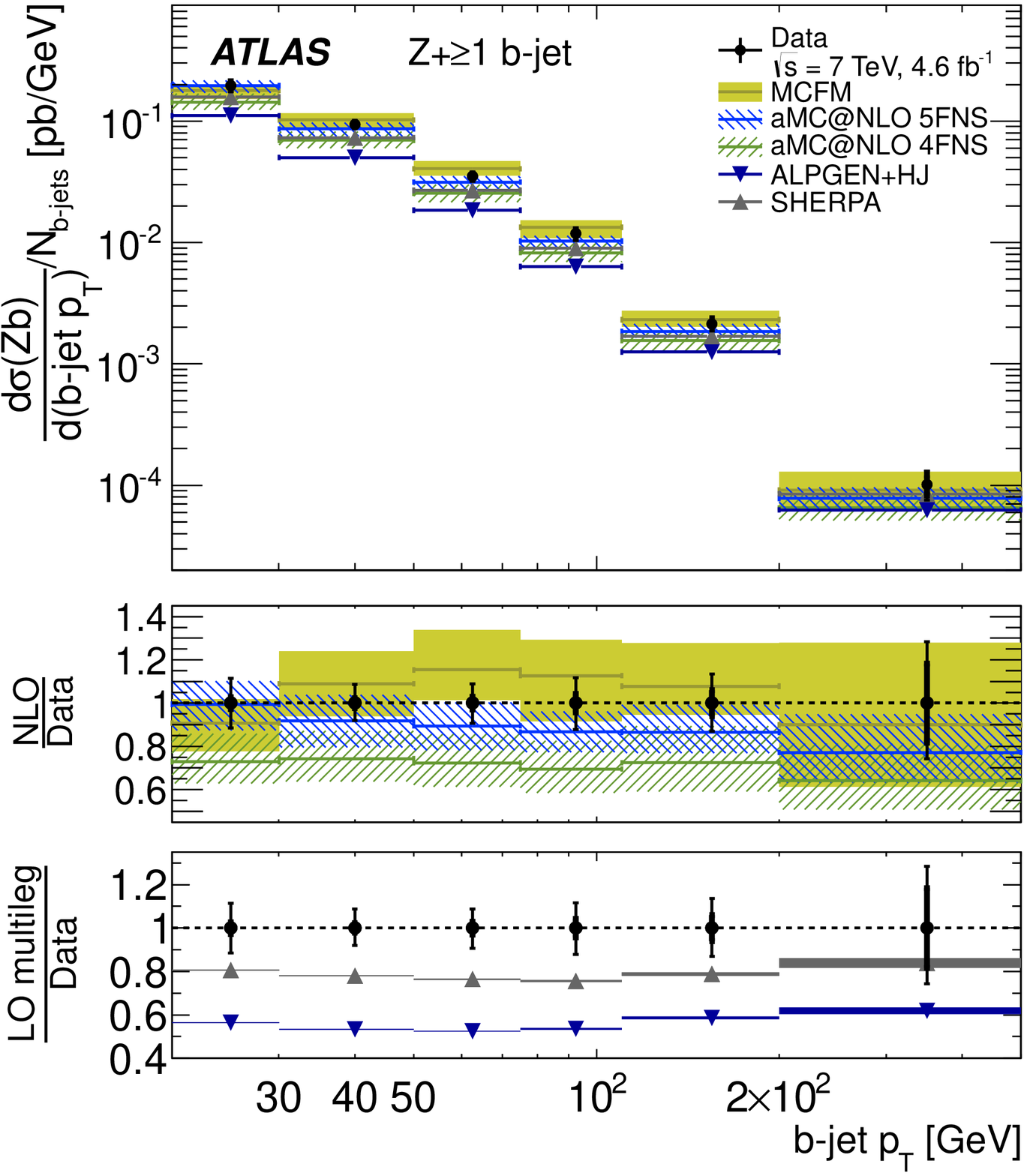}
  \label{fig:results_bjet_pT}
}
\subfigure[ ]{
  \includegraphics[width=0.5\textwidth]{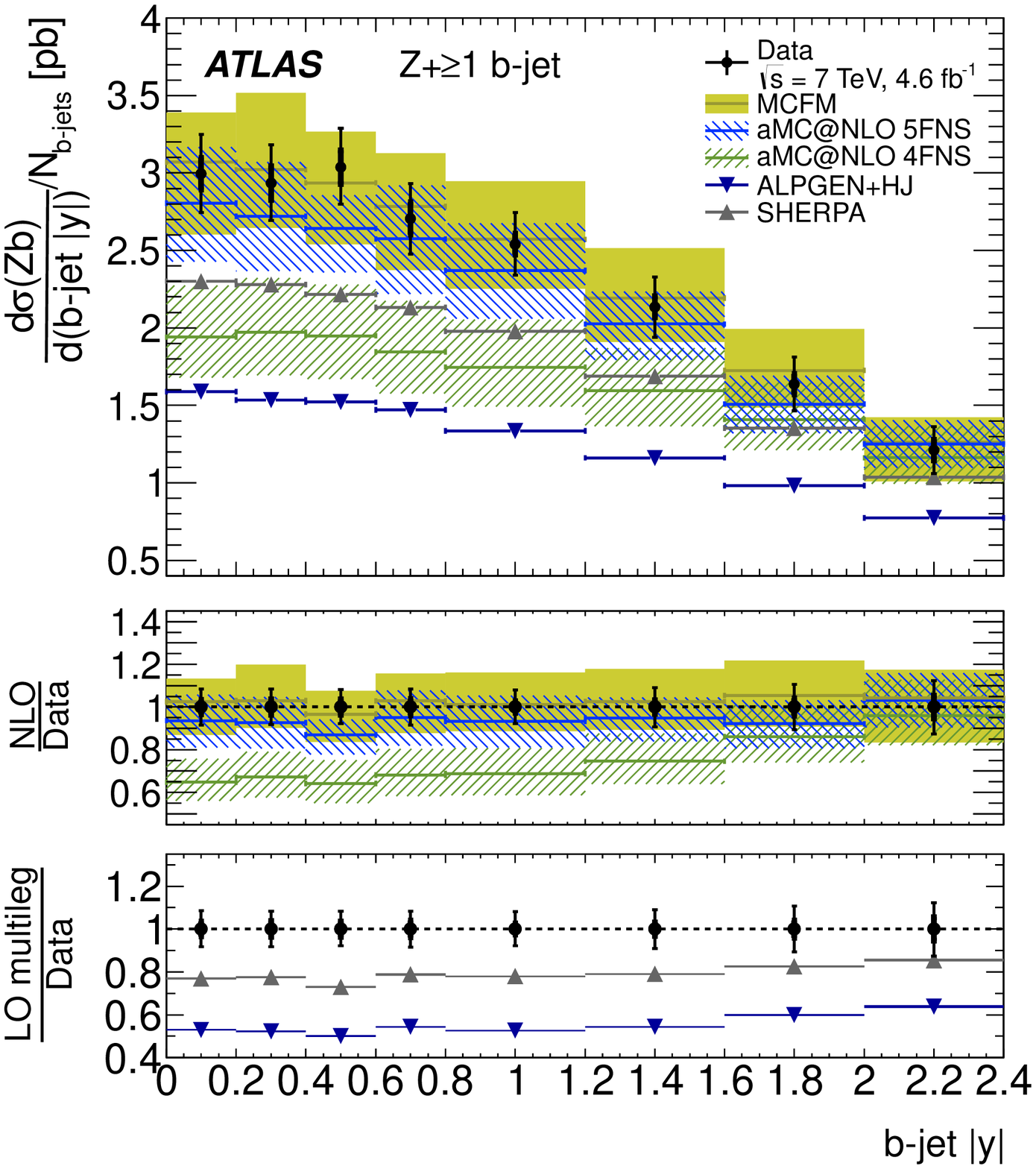}
  \label{fig:results_bjet_y}
}
\end{center}
\caption{The inclusive $b$-jet cross-section \sigZbjet\ as a function of  $b$-jet \pt\ (a) and \absy\ (b). 
The top panels show measured differential cross-sections as filled circles with statistical (inner) and 
total (outer bar) uncertainties. Overlayed for comparison are the NLO predictions from \mcfm\ and \amcnlo\ both using the MSTW2008 PDF set. 
The shaded bands represents the total theoretical uncertainty for \mcfm\ and the uncertainty 
bands on \amcnlo\ points represent the dominant theoretical scale uncertainty only. Also overlaid are LO multi-legged predictions for 
\alpgen+\herwig+\jimmy\ and \sherpa. The middle panels show the ratio of NLO predictions to data, and the lower panels show the ratio of LO
predictions to data.
\label{fig:results_bjet_pTy}}
\end{figure}

\begin{figure}[p]
\begin{center}
\subfigure[ ]{
  \includegraphics[width=0.5\textwidth]{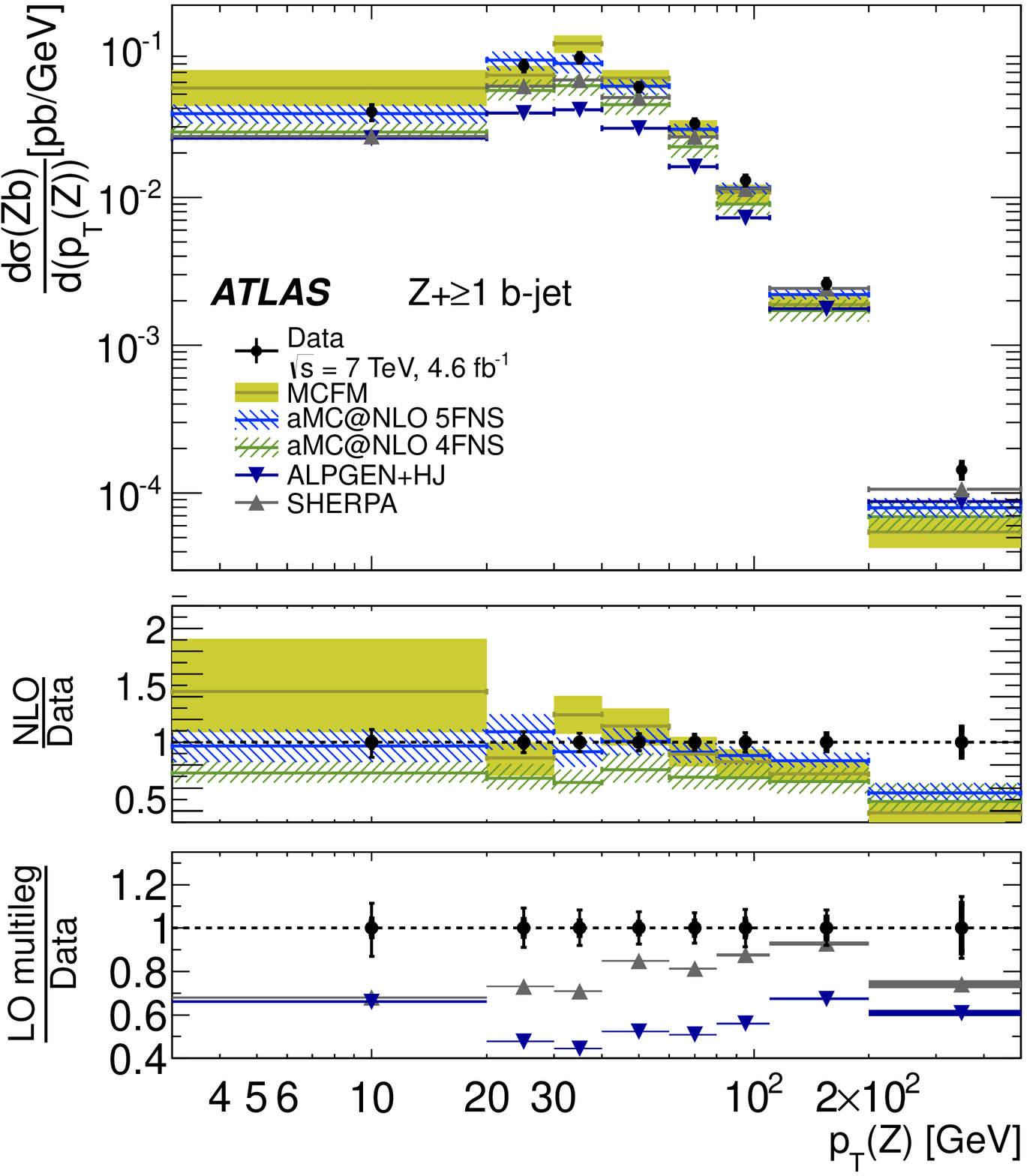}
  \label{fig:results_Z_pT}
}
\subfigure[ ]{
  \includegraphics[width= 0.5\textwidth]{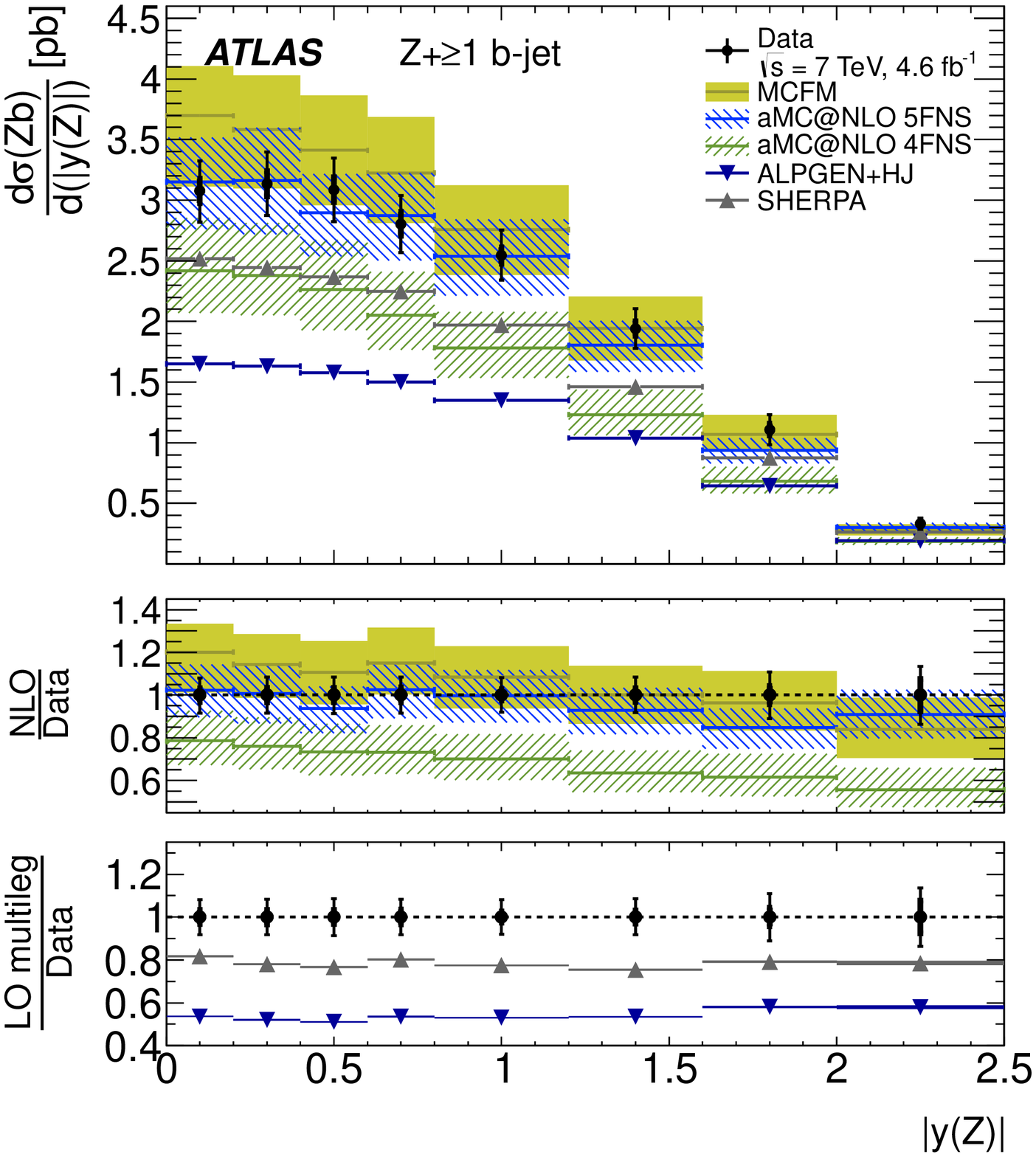}
  \label{fig:results_Z_y}
}
\end{center}
\caption{The cross-section \sigZb\ as a function of $Z$ boson \pt\ (a) and \absy\ (b). 
The top panels show measured differential cross-sections as filled circles with statistical (inner) and 
total (outer bar) uncertainties. Overlayed for comparison are the NLO predictions from \mcfm\ and \amcnlo\ both using the MSTW2008 PDF set.
The shaded bands represents the total theoretical uncertainty for \mcfm\ and the uncertainty 
bands on \amcnlo\ points represent the dominant theoretical scale uncertainty only. Also overlaid are LO multi-legged predictions for 
\alpgen+\herwig+\jimmy\ and \sherpa. The middle panels show the ratio of NLO predictions to data, and the lower panels show the ratio of LO
predictions to data.
\label{fig:results_Z_pTy}}
\end{figure}

In general, good agreement with the data can be seen for $\Delta y(Z,b)$ and $y_{\mathrm{boost}}(Z,b)$ in Figure~\ref{fig:results_bjet_angles}, though with some evidence for a slope in the ratio of \amcnlo\ 4FNS relative to the data for $y_{\mathrm{boost}}(Z,b)$.
In $\Delta\phi(Z,b)$ (Figure~\ref{fig:results_bjet_angles_20}) the fixed-order pQCD prediction from \mcfm\ has significant discrepancy at $\Delta\phi(Z,b)=\pi$, which also distorts the $\Delta R(Z,b)$\ prediction. 
This is due to the fixed-order calculation containing at most one or two outgoing partons in association with the $Z$ boson. 
In the case of one parton, $\Delta\phi(Z,b) = \pi$ by construction, leading to the distorted distribution.
The inclusion of higher multiplicity matrix elements in \alpgen\ and \sherpa, and matching to parton 
shower models in \alpgen, \sherpa\ and \amcnlo\ helps to populate the $\Delta\phi(Z,b)$ distribution in a way
which yields a much better agreement with data. This emphasises the importance of higher order effects when considering such 
distributions.
The region of low $\Delta\phi(Z,b)$, which is most sensitive to additional QCD radiation as well as soft 
corrections, is also poorly modelled by \mcfm; these effects are not fully captured in the non-perturbative corrections applied to that prediction.

\begin{figure}[p]
\begin{center}
\subfigure[ ]{
  \includegraphics[width=0.5\textwidth]{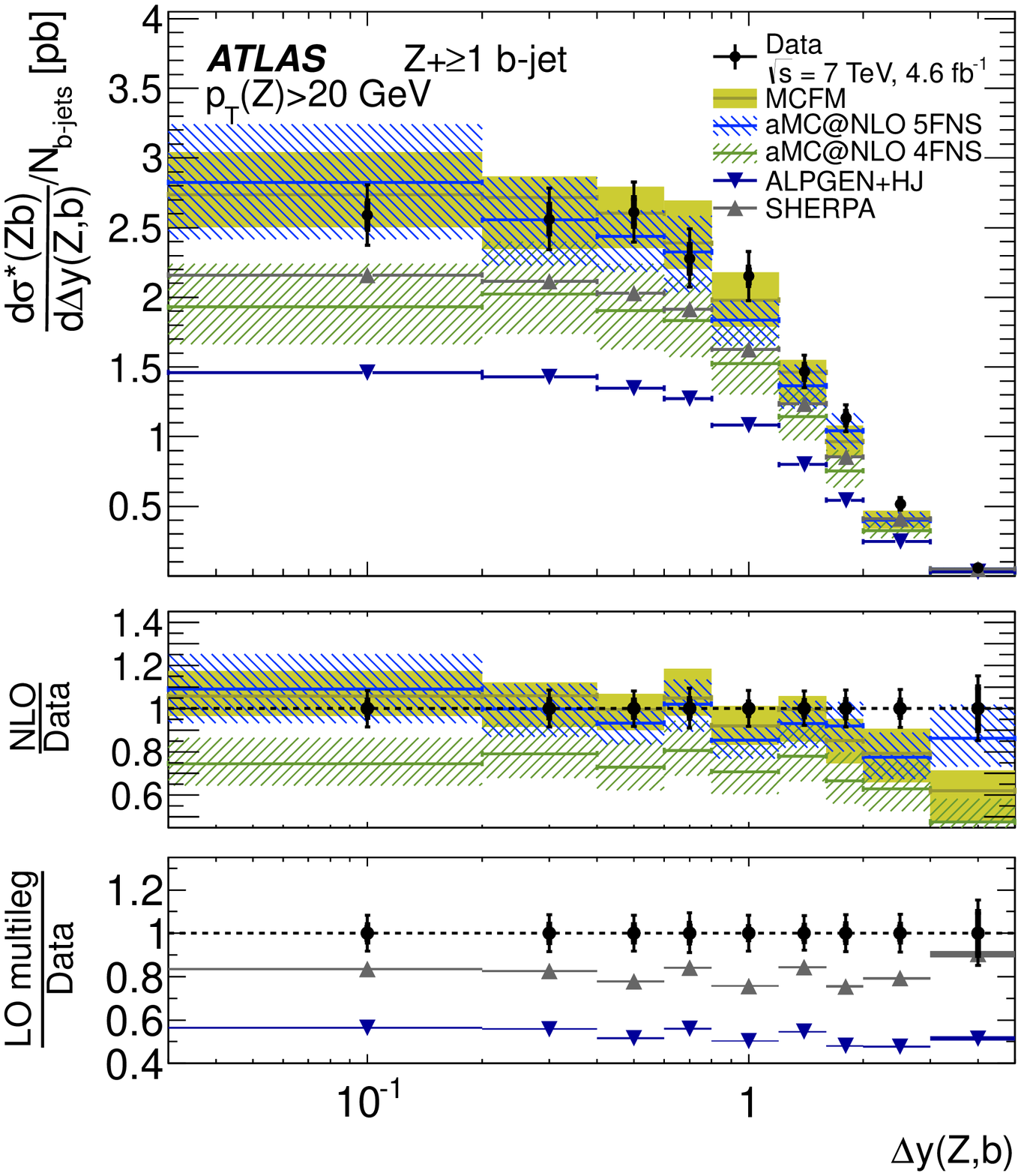}
  \label{fig:results_bjet_dy}
}
\subfigure[ ]{
  \includegraphics[width= 0.5\textwidth]{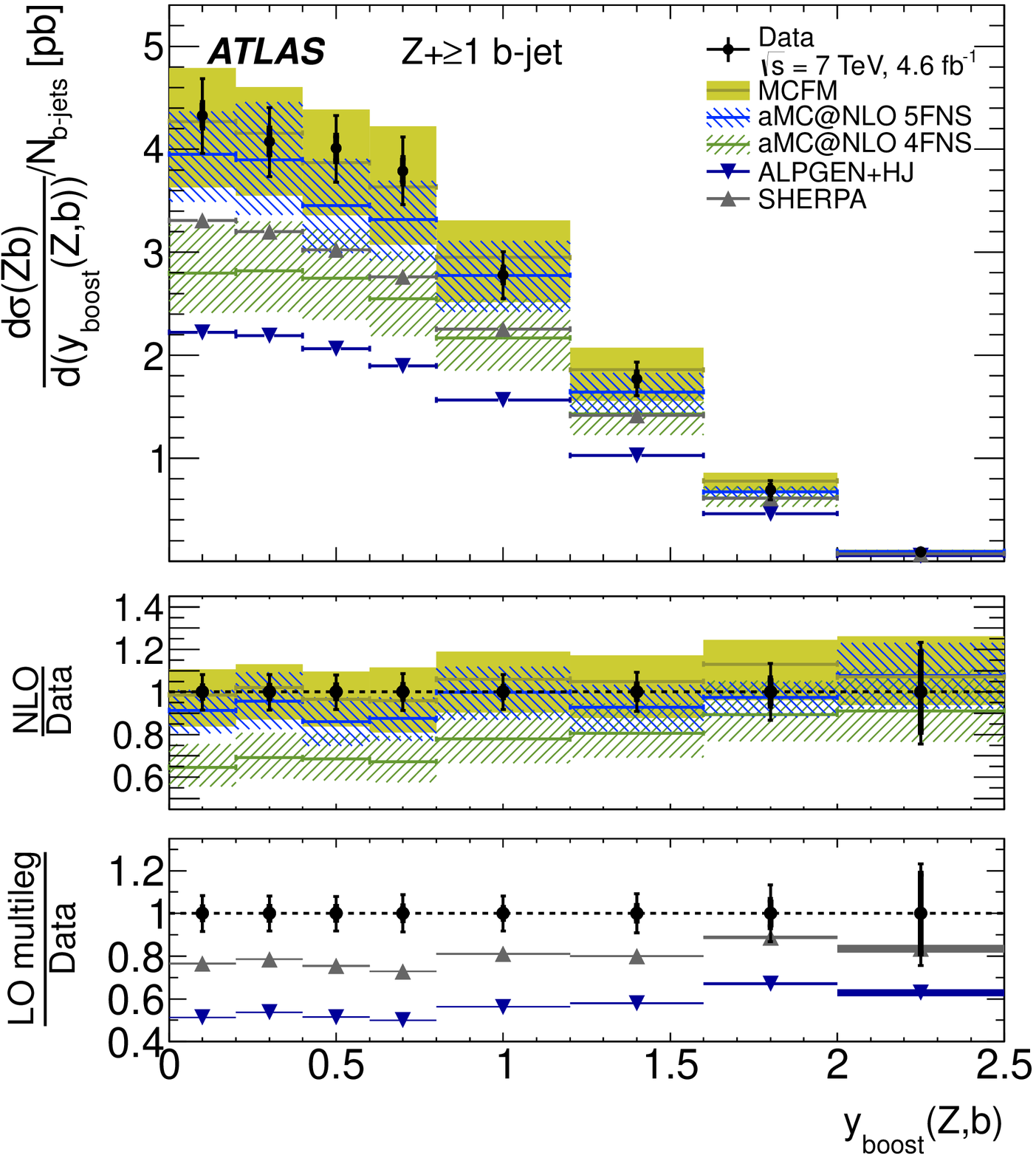}
  \label{fig:results_bjet_yb}
}
\end{center}
\caption{The inclusive $b$-jet cross-sections \sigZbjetstar\ as a function of $\Delta y(Z,b)$ (a) and \sigZbjet\ as a function of $y_{\mathrm{boost}}(Z,b)$ (b). The
former inclusive cross-section requires that the $Z$ boson \pt\ be at least 20~\gev.
The top panels show measured differential cross-sections as filled circles with statistical (inner) and 
total (outer bar) uncertainties. Overlayed for comparison are the NLO predictions from \mcfm\ and \amcnlo\ both using the MSTW2008 PDF set.
The shaded bands represents the total theoretical uncertainty for \mcfm\ and the uncertainty 
bands on \amcnlo\ points represent the dominant theoretical scale uncertainty only. Also overlaid are LO multi-legged predictions for 
\alpgen+\herwig+\jimmy\ and \sherpa. The middle panels show the ratio of NLO predictions to data, and the lower panels show the ratio of LO
predictions to data. \label{fig:results_bjet_angles}}
\end{figure}

\begin{figure}[p]
\begin{center}
\subfigure[ ]{
  \includegraphics[width=0.5\textwidth]{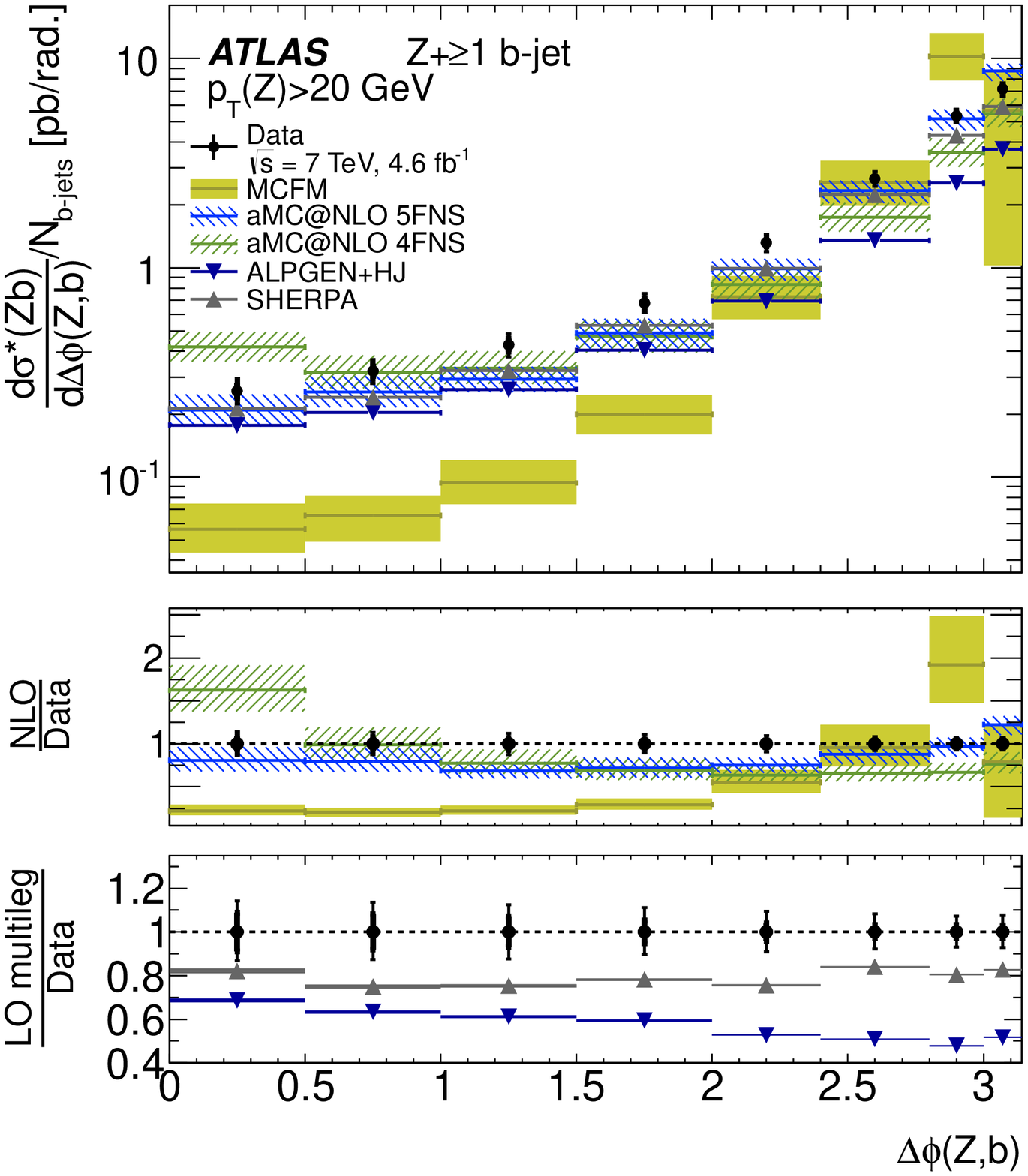}
  \label{fig:results_bjet_angles_20_dphi}
}
\subfigure[ ]{
  \includegraphics[width= 0.5\textwidth]{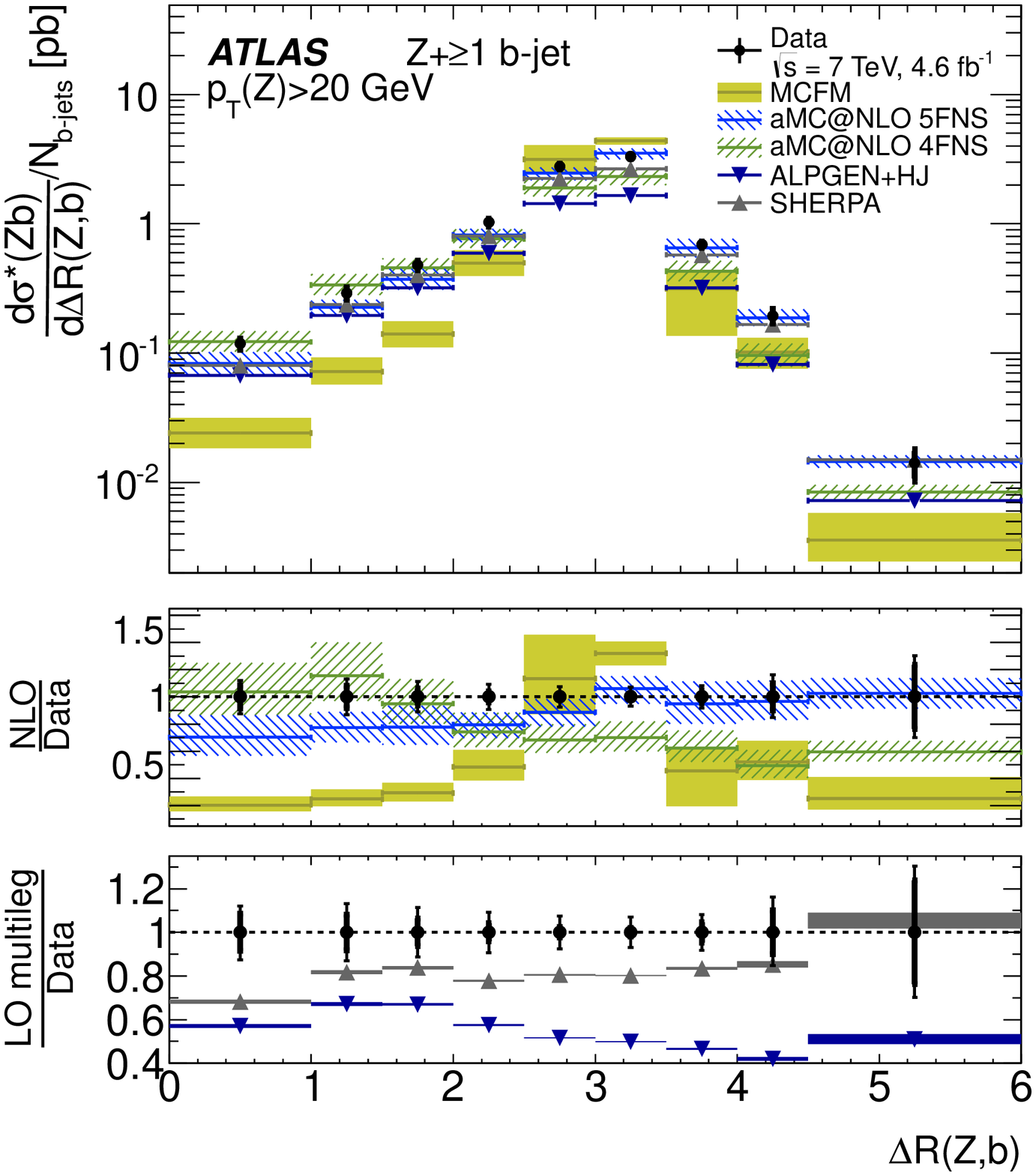}
  \label{fig:results_bjet_angles_20_dR}
}
\end{center}
\caption{The inclusive $b$-jet cross-section \sigZbjetstar\ as a function of $\Delta \phi(Z,b)$ (a) and \dRZb\ (b). The inclusive cross-section requires that the $Z$ boson \pt\ be at least 20~\gev.
The top panels show measured differential cross-sections as filled circles with statistical (inner) and 
total (outer bar) uncertainties. Overlayed for comparison are the NLO predictions from \mcfm\ and \amcnlo\ both using the MSTW2008 PDF set.
The shaded bands represents the total theoretical uncertainty for \mcfm\ and the uncertainty 
bands on \amcnlo\ points represent the dominant theoretical scale uncertainty only. Also overlaid are LO multi-legged predictions for 
\alpgen+\herwig+\jimmy\ and \sherpa. The middle panels show the ratio of NLO predictions to data, and the lower panels show the ratio of LO
predictions to data.\label{fig:results_bjet_angles_20}}
\end{figure}

For the  \Zbb\ differential cross-sections shown in Figures~\ref{fig:results_Zbb_Z} and \ref{fig:results_Zbb_bjet}, all predictions provide reasonable descriptions of the data within the large experimental uncertainties. 
There is some evidence for disagreements between predictions and data at low $m(b,b)$ and low $\Delta R(b,b)$. 

\begin{figure}[p]
\begin{center}
\subfigure[ ]{
  \includegraphics[width=0.5\textwidth]{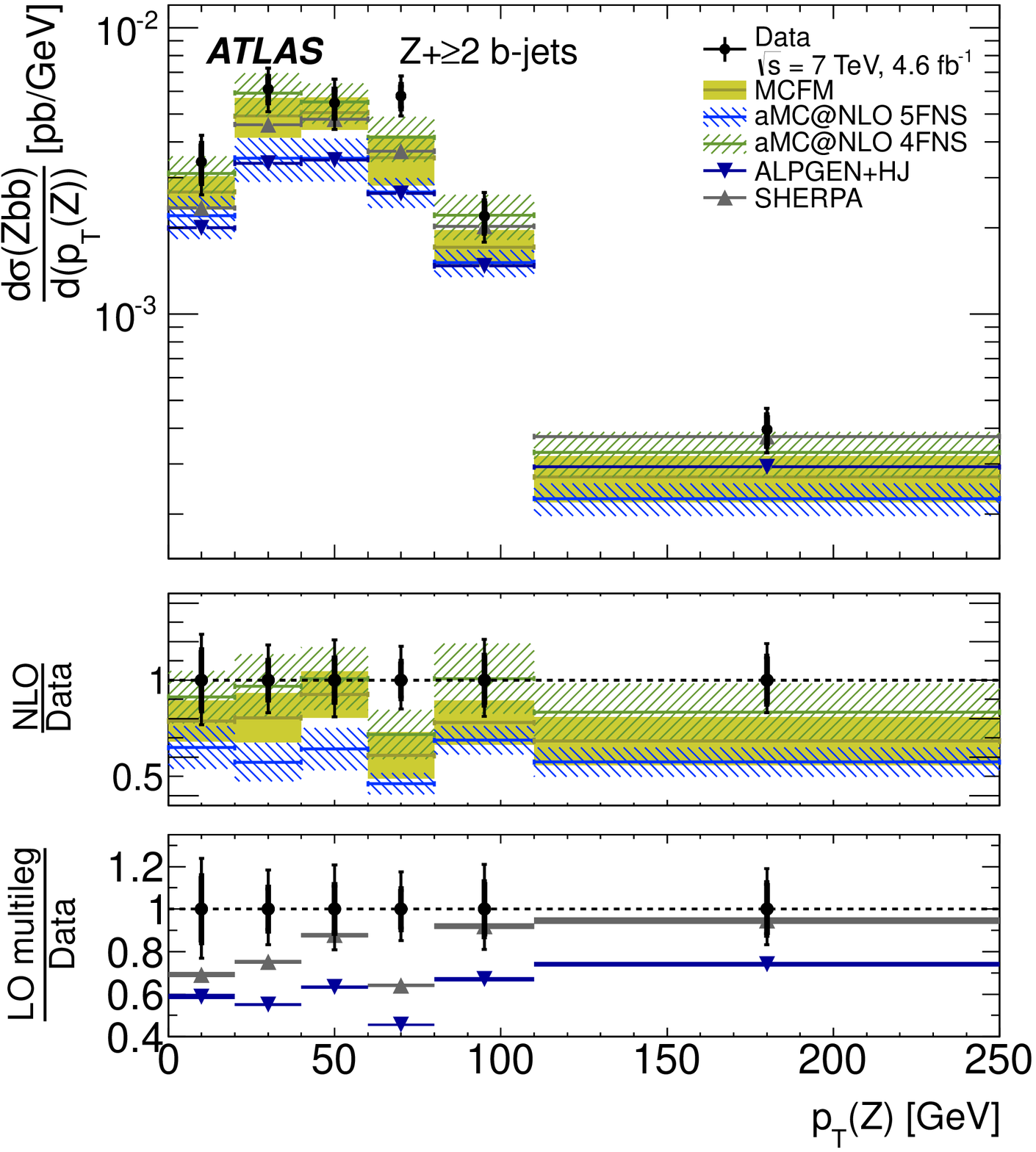}
  \label{fig:results_Zbb_ZpT}
}
\subfigure[ ]{
  \includegraphics[width= 0.5\textwidth]{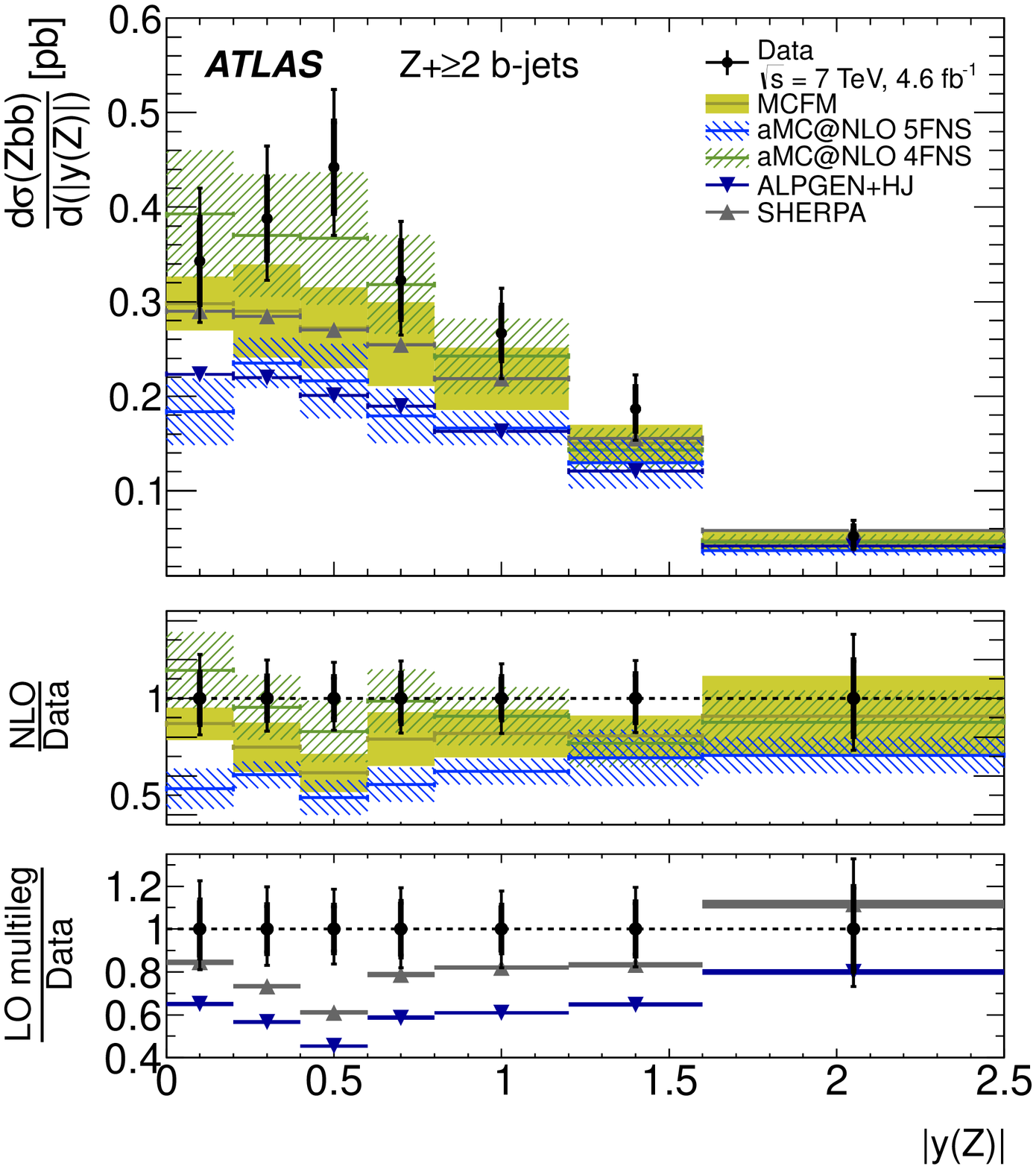}
  \label{fig:results_Zbb_Zy}
}
\end{center}
\caption{The cross-section \sigZbb\ as a function of  $Z$ boson \pt\ (a), and \absy\ (b).
The top panels show measured differential cross-sections as filled circles with statistical (inner) and 
total (outer bar) uncertainties. Overlayed for comparison are the NLO predictions from \mcfm\ and \amcnlo\ both using the MSTW2008 PDF set.
The shaded bands represents the total theoretical uncertainty for \mcfm\ and the uncertainty 
bands on \amcnlo\ points represent the dominant theoretical scale uncertainty only. Also overlaid are LO multi-legged predictions for 
\alpgen+\herwig+\jimmy\ and \sherpa. The middle panels show the ratio of NLO predictions to data, and the lower panels show the ratio of LO
predictions to data.\label{fig:results_Zbb_Z}}
\end{figure}

\begin{figure}[p]
\begin{center}
\subfigure[ ]{
  \includegraphics[width= 0.5\textwidth]{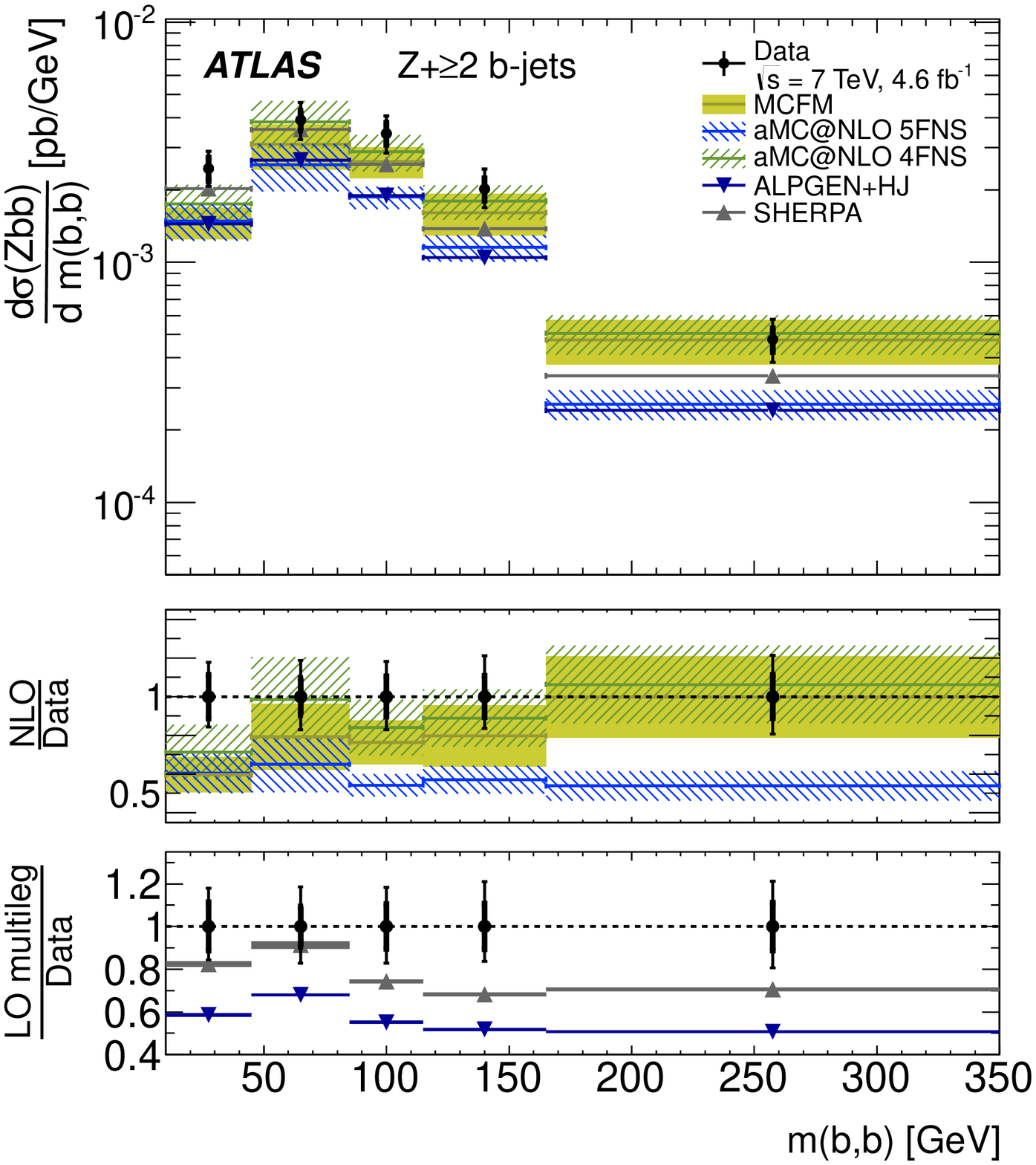}
  \label{fig:results_Zbb_bbmass}
}
\subfigure[ ]{
  \includegraphics[width=0.5\textwidth]{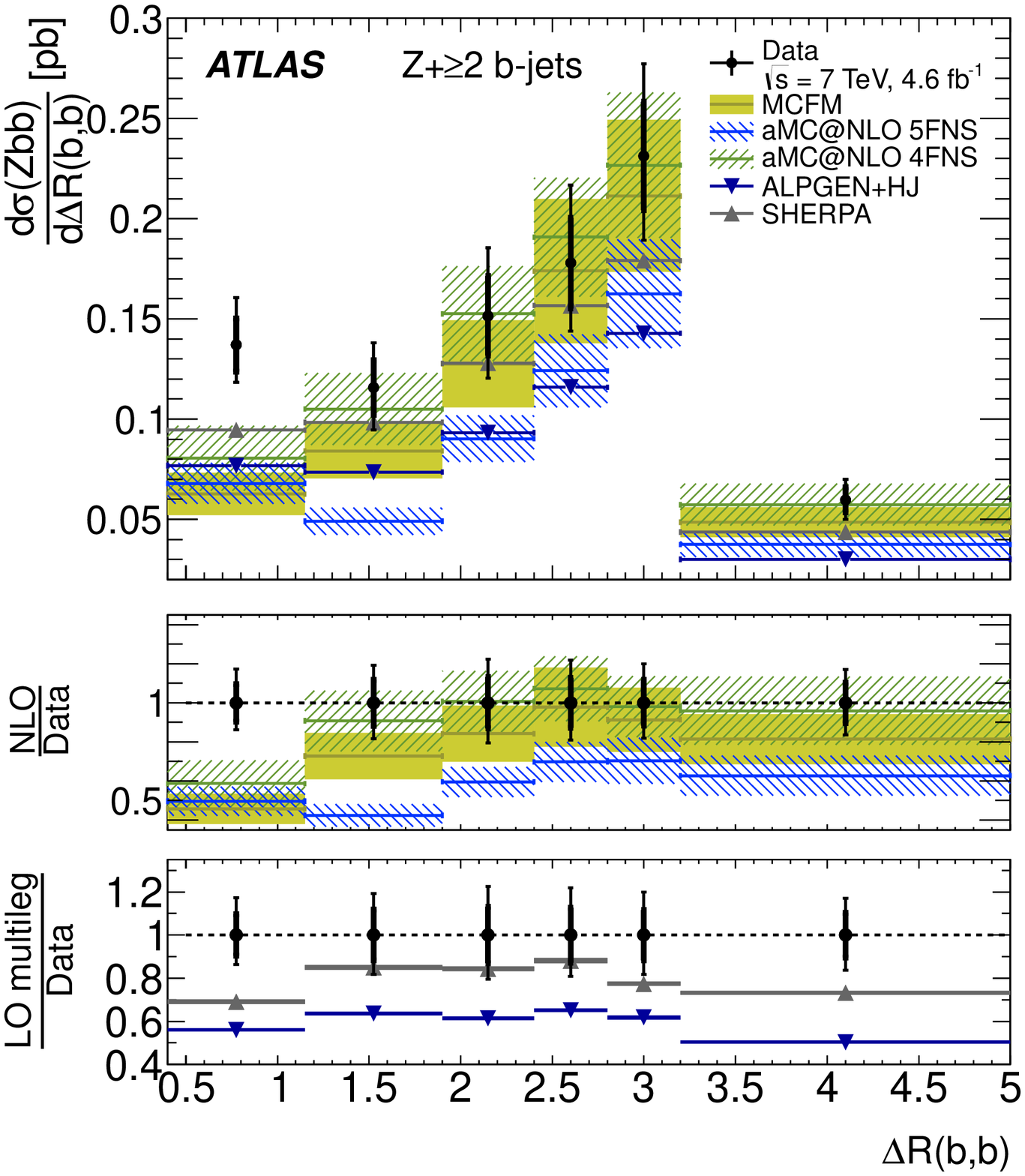}
  \label{fig:results_Zbb_bbdR}
}
\end{center}
\caption{The cross-section \sigZbb\ as a function of \mbb\ (a) and \dRbb\ (b).
The top panels show measured differential cross-sections as filled circles with statistical (inner) and 
total (outer bar) uncertainties. Overlayed for comparison are the NLO predictions from \mcfm\ and \amcnlo\ both using the MSTW2008 PDF set.
The shaded bands represents the total theoretical uncertainty for \mcfm\ and the uncertainty 
bands on \amcnlo\ points represent the dominant theoretical scale uncertainty only. Also overlaid are LO multi-legged predictions for 
\alpgen+\herwig+\jimmy\ and \sherpa. The middle panels show the ratio of NLO predictions to data, and the lower panels show the ratio of LO
predictions to data.\label{fig:results_Zbb_bjet}}
\end{figure}

Finally, Figure~\ref{fig:results_Zy_pdf} compares the \mcfm\ predictions obtained using different PDFs
to the data for the $Z$ boson rapidity distribution, which is the distribution found to have the largest dependence on the
PDF set used.
It can be seen that, while the different PDF sets do yield different results, they all show a 
similar trend relative to the data, and the differences are small compared to the theoretical scale uncertainty.

\begin{figure}[p]
\begin{center}
\subfigure[ ]{
  \includegraphics[width=0.6\textwidth]{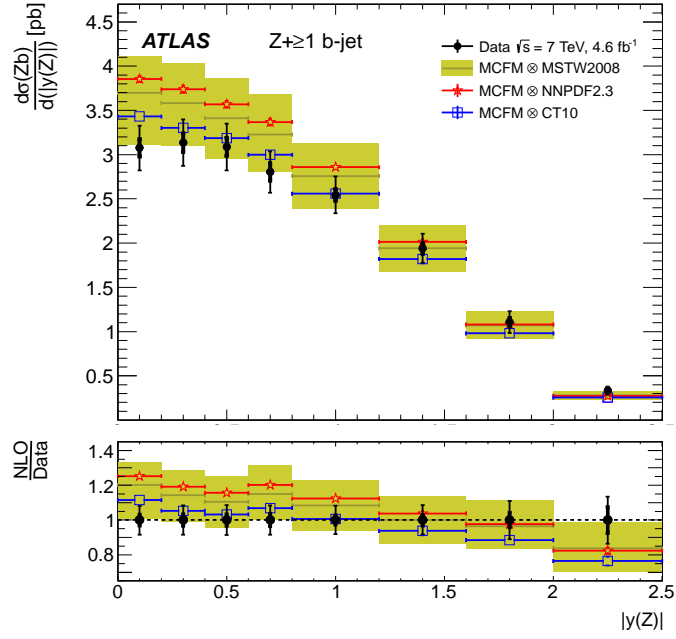}
  \label{fig:results_Zy_b_pdf}
}
\subfigure[ ]{
  \includegraphics[width= 0.6\textwidth]{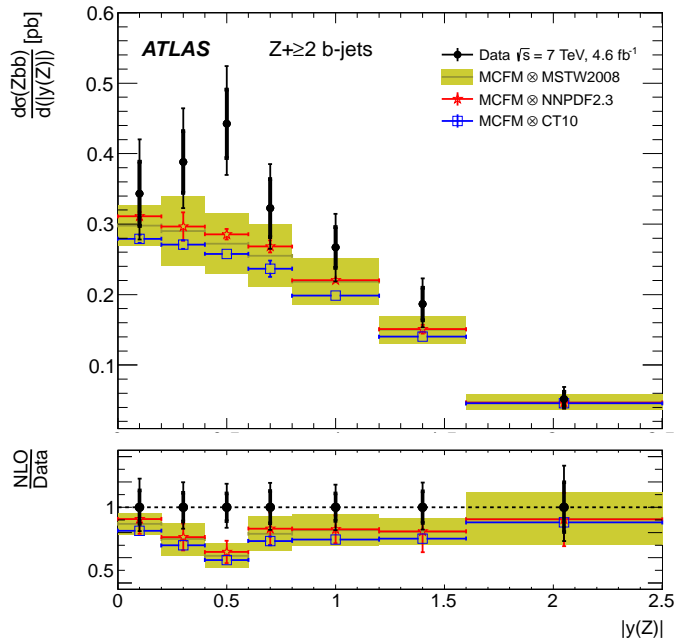}
  \label{fig:results_Zy_bb_pdf}
}
\end{center}
\caption{The \mcfm\ prediction using different PDF sets for the cross-sections \sigZb\ (a) and \sigZbb\ (b) as a function of the $Z$ boson \absy. 
The top panels show measured differential cross-sections as filled circles with 
 statistical (inner) and total (outer bar) uncertainties. 
The shaded band represents the total theoretical uncertainty for \mcfm\ interfaced to the MSTW2008 PDF set.
Uncertainties on \mcfm\ predictions with alternative PDF sets are statistical only.
The lower panel shows the ratio of each prediction to data. \label{fig:results_Zy_pdf}}
\end{figure}

\clearpage

\section{Conclusions}
\label{sec:Conclusions}

Differential $Z$+$b$-jets cross-section measurements from the LHC have been presented using 4.6~\ifb\ of 
\rts=7~\tev\ $pp$ collision data recorded by the ATLAS detector in 2011. In total, 12 
distributions for \Zb\ and \Zbb\ topologies have been investigated and compared to theoretical pQCD calculations. 
Next-to-leading-order predictions from \mcfm\ 
and \amcnlo\ generally provide the best overall description of the data. The agreement 
of the \amcnlo\ cross-section prediction with data differs in the \Zb\ and \Zbb\ cases, with the former better described by the 5FNS prediction
and the latter better described by the 4FNS prediction.
Even at NLO, scale uncertainties dominate and currently limit any sensitivity to different PDF sets.
Descriptions of the shapes of the differential cross-sections are generally good within uncertainties for both LO and NLO predictions.
For angular distributions in the \Zb\ selection, where the fixed-order NLO prediction
is observed to break down, the differential shapes in data are well modelled by LO multi-legged predictions.




\section{Acknowledgements}

We thank CERN for the very successful operation of the LHC, as well as the
support staff from our institutions without whom ATLAS could not be
operated efficiently.

We acknowledge the support of ANPCyT, Argentina; YerPhI, Armenia; ARC,
Australia; BMWFW and FWF, Austria; ANAS, Azerbaijan; SSTC, Belarus; CNPq and FAPESP,
Brazil; NSERC, NRC and CFI, Canada; CERN; CONICYT, Chile; CAS, MOST and NSFC,
China; COLCIENCIAS, Colombia; MSMT CR, MPO CR and VSC CR, Czech Republic;
DNRF, DNSRC and Lundbeck Foundation, Denmark; EPLANET, ERC and NSRF, European Union;
IN2P3-CNRS, CEA-DSM/IRFU, France; GNSF, Georgia; BMBF, DFG, HGF, MPG and AvH
Foundation, Germany; GSRT and NSRF, Greece; ISF, MINERVA, GIF, I-CORE and Benoziyo Center,
Israel; INFN, Italy; MEXT and JSPS, Japan; CNRST, Morocco; FOM and NWO,
Netherlands; BRF and RCN, Norway; MNiSW and NCN, Poland; GRICES and FCT, Portugal; MNE/IFA, Romania; MES of Russia and ROSATOM, Russian Federation; JINR; MSTD,
Serbia; MSSR, Slovakia; ARRS and MIZ\v{S}, Slovenia; DST/NRF, South Africa;
MINECO, Spain; SRC and Wallenberg Foundation, Sweden; SER, SNSF and Cantons of
Bern and Geneva, Switzerland; NSC, Taiwan; TAEK, Turkey; STFC, the Royal
Society and Leverhulme Trust, United Kingdom; DOE and NSF, United States of
America.

The crucial computing support from all WLCG partners is acknowledged
gratefully, in particular from CERN and the ATLAS Tier-1 facilities at
TRIUMF (Canada), NDGF (Denmark, Norway, Sweden), CC-IN2P3 (France),
KIT/GridKA (Germany), INFN-CNAF (Italy), NL-T1 (Netherlands), PIC (Spain),
ASGC (Taiwan), RAL (UK) and BNL (USA) and in the Tier-2 facilities
worldwide.

%
\bibliographystyle{atlasBibStyleWithTitle}
\bibliography{ZbZbbPaper}

\onecolumn
\clearpage
\begin{flushleft}
{\Large The ATLAS Collaboration}

\bigskip

G.~Aad$^{\rm 84}$,
B.~Abbott$^{\rm 112}$,
J.~Abdallah$^{\rm 152}$,
S.~Abdel~Khalek$^{\rm 116}$,
O.~Abdinov$^{\rm 11}$,
R.~Aben$^{\rm 106}$,
B.~Abi$^{\rm 113}$,
M.~Abolins$^{\rm 89}$,
O.S.~AbouZeid$^{\rm 159}$,
H.~Abramowicz$^{\rm 154}$
% ...
\bigskip
\\
% ...
$^{11}$ Institute of Physics, Azerbaijan Academy of Sciences, Baku, Azerbaijan\\
% ...
$^{84}$ CPPM, Aix-Marseille Universit{\'e} and CNRS/IN2P3, Marseille, France\\
% ...
$^{89}$ Department of Physics and Astronomy, Michigan State University, East Lansing MI, United States of America\\
% ...
$^{106}$ Nikhef National Institute for Subatomic Physics and University of Amsterdam, Amsterdam, Netherlands\\
% ...
$^{112}$ Homer L. Dodge Department of Physics and Astronomy, University of Oklahoma, Norman OK, United States of America\\
$^{113}$ Department of Physics, Oklahoma State University, Stillwater OK, United States of America\\
% ...
$^{116}$ LAL, Universit{\'e} Paris-Sud and CNRS/IN2P3, Orsay, France\\
% ...
$^{152}$ Institute of Physics, Academia Sinica, Taipei, Taiwan\\
% ...
$^{154}$ Raymond and Beverly Sackler School of Physics and Astronomy, Tel Aviv University, Tel Aviv, Israel\\
% ...
$^{159}$ Department of Physics, University of Toronto, Toronto ON, Canada\\
% ...
\end{flushleft}


\end{document}